\documentclass[%
 preprint,
 12pt,
 amsmath,amssymb,
 aps,
]{iopart}

\usepackage{hyperref}
\usepackage{subfigure}
\usepackage{xparse}
\usepackage{xcolor}
\usepackage{color}
\usepackage{multirow}
\usepackage[utf8]{inputenc}
\usepackage[T1]{fontenc}

\newcommand{\scl}{0.40}

\usepackage{graphicx}
\usepackage{dcolumn}
\usepackage{bm}

\begin{document}


\title{Particle momentum variation under interaction with wavepackets of finite spatial width}

\author{Theodoros Bournelis$^1$ and Yannis Kominis$^1$}

\address{%
$^1$School of Applied Mathematical and Physical Sciences,\\ National Technical University of Athens, Athens 15780, Greece
}%
\vspace{10pt}
\ead{gkomin@central.ntua.gr}
\vspace{10pt}
\begin{indented}
\item[]
\end{indented}

\date{\today}

\begin{abstract}
   The single and collective particle interaction with spatially localized wavepackets is analytically and numerically studied. The role of the finite spatial width of the wavepacket on the momentum and energy variation of particles passing through the wavepacket is investigated. The range of validity of analytical results, based on a perturbative approach, is investigated and clearly defined. Strongly nonlinear effects are shown to qualitatively differentiate the collective particle dynamics, for larger interaction strengths. These effects are manifested through the complex dependency of averaged momentum variations on the initial particle momentum, as shown by dissecting a particle distribution in terms of partitioning the ensemble of particles with respect to their initial momentum. The results provide understanding on the relation between single and collective particle dynamics and the emerging of complexity for weak and strong particle interactions with localized wavepackets.
\end{abstract}

\maketitle


\section{\label{sec:level1}Introduction}

Wave-particle interaction is a ubiquitous mechanism under which momentum and energy is exchanged between waves and particles. In fusion plasmas applications, external high-frequency waves are used in order to heat plasmas as well as to drive current by imparting energy and momentum to particles through ion-cyclotron or electron-cyclotron resonances, whereas low-frequency waves corresponding to plasma instabilities result in particle energy and momentum transport of either thermal or fast particles, with crucial consequences on plasma confinement \cite{Freidberg_book, Miyamoto_book}. In space plasmas, Langmuir waves are known to occur in the solar wind and to be correlated with the presence of suprathermal electron beams \cite{Rozmus_88b, Zaslavsky_10}. In all cases, particles do not interact with monochromatic (plane) waves but with wavepackets having a finite spectral and spatial width, that form potential barriers appearing in many applications of wave-particle interactions \cite{Dodin_06}.   

The wave-particle interaction is an intrinsically complex nonlinear phenomenon, that has been widely investigated for more than four decades. Single particle dynamics under the presence of monochromatic electrostatic waves has been studied as a fundamental example of Hamiltonian chaos \cite{Smith_75, Taylor_75, Fukuyama_77, Karney_77, Doveil_81, Ram_86, Skiff_87, Karimabadi_90, Strozzi_03, Benisti_07, Jorns_11, Escande_16, Lichtenberg_book}. The case of spatially modulated waves having the form of localized wavepackets, has also been studied in terms of single \cite{Pocobelli_81, Menyuk_85, Fuchs_85, Rozmus_88,  Bruhwiler_92, Farina_91, Akimoto_97, Akimoto_02, Artemyev_19a, Soni_21} and collective particle dynamics \cite{Denavit_72, Morales_74, Bezzerides_75, Fuchs_85, Rozmus_88, Artemyev_19a, Artemyev_19b}, when the spatial width of the envelope is relatively large in comparison to the wavelength of the carrier wave. Studies on single particle dynamics are based in phase space analysis with the utilization of Poincare surfaces of section whereas studies on the collective particle dynamics consider Fokker-Planck or Vlasov equation for the evolution of the particle momentum distribution function under interaction with a wavepacket. The application of adiabatic and perturbation methods has provided analytical results either for slowly varying \cite{Pocobelli_81, Menyuk_85, Farina_91, Bruhwiler_92} or for small-amplitude \cite{Kominis_06, Kominis_08, Kominis_10, Kominis_12} wavepackets. However, the role of the finite spatial width of a wavepacket on the single particle dynamics as well as on the collective momentum variation of particle ensembles having various initial momentum distributions is still a subject of research interest, due to its relevance to realistic applications where the aforementioned simplifications might not be applicable.

In this work, in order to focus on the role of the finite spatial width of a wavepacket on the particle momentum variation, we consider the case of an electrostatic wavepacket of arbitrary group/phase velocity, amplitude, and a width ranging from few-cycle to adiabatic wavepackets. A zero background magnetic field is considered. In this case, particle dynamics are equivalent with the case of a non-zero uniform magnetic field with a wavepacket propagating along its direction, in terms of parallel particle dynamics, since dynamics in the directions parallel and perpendicular to the magnetic field actually decouple. The electrostatic waves correspond to Langmuir or ion acoustic waves propagating either in an unmagnetized plasma, or along the direction of a background magnetic field. Similar configurations have also been considered for the study of heating and current drive with lower-hybrid waves in fusion plasmas \cite{Fuchs_85, Bruhwiler_92} with the group velocity of the wavepacket having a zero component along the direction of the phase velocity. The above simplifications do not essentially affect the role of the finite spatial width on the wave-particle interaction and result in a one-dimensional configuration (one-dimensional electrostatic wave with parallel group and phase velocities) that facilitates the systematic numerical investigation of the particle momentum variation for arbitrary wavepacket parameters.     

The main aim of this work is twofold: to systematically explore the range of validity of analytical results based on perturbation theory, and to identify the emergence of stronger complexity of particle dynamics in the range where the perturbation theory is no longer valid. In doing so, we adopt an approach according to which the collective dynamics of an ensemble of particles is dissected in terms of their initial momenta, in order to relate the single particle dynamics with the collective behavior, as expressed with certain momentum averages over initial conditions. This approach provides understanding on the contribution of specific particle "beamlets" of a certain initial momentum on the modification and the final form of the total particle momentum distribution after interaction with the wavepacket. For increasing perturbation strength, the quantitative differences between analytical results eventually evolve to significant qualitative differences, due to complex dynamical features that cannot be captured by the perturbation theory.    

Analytical results for the collective particle dynamics are presented for the limiting cases of either small-amplitude or infinite-width wavepackets and their domains of validity are identified in the parameter space of the wavepacket by systematic comparison with numerical results. It is worth emphasizing that such a validation is enabled by the one-dimensional character of the model, since collective behaviour can be simulated by a smaller number of particles in comparison to higher-dimensional configurations where the number of particles required to fill higher-dimensional phase spaces results in a significant computational cost. 
Beyond the range of validity of the analytical results, the strongly nonlinear character of the interaction is exhibited through the complex dependence of the certain momentum variation averages on the initial particle momentum as well as through the purely nonlinear effect of a non-zero mean momentum variation for a uniform particle momentum distribution function beyond the standard quasi-linear effect of the Landau resonance \cite{Dawson_61, Drummond_04}. 
The results of this study provide insight regarding the qualitative characteristics of higher-dimensional configurations and/or interactions with electromagnetic wavepackets of finite spatial width, as well as the domain of validity of analytical results, in such cases.

\section{Hamiltonian Model and Analytical Results}

The Hamiltonian describing charged particle motion under the presence of a spatially localized electrostatic wavepacket is
\begin{equation}
    H(p,z,t)=\frac{p^2}{2}+\Phi(z-v_g t)\sin\left(k(z-v_pt)\right)
\end{equation}
where $p,z$ are the particle momentum and position, the particle mass and charge have been set equal to unity, $k$ is the wavenumber of the carrier wave, $v_g, v_p$ are the group and phase velocities, and $\Phi$ is the spatial profile of the wavepacket. 
A canonical transformation to the group velocity frame is given by the mixed variable generating function
\begin{equation}
    F_2(z,\bar{p},t)=(z-v_gt)(\bar{p}+v_g)
\end{equation}
where the new (barred) variables are related to the original ones as
\begin{eqnarray}
    p&=&\partial F_2 / \partial z = \bar{p}+v_g \nonumber \\
    \bar{z}&=&\partial F_2 / \partial \bar{p} = z-v_g t
\end{eqnarray}
and the new Hamiltonian becomes
\begin{equation}
    \bar{H}(\bar{p},\bar{z},t)=\frac{\bar{p}^2}{2}+\Phi(\bar{z})\sin\left(k(\bar{z}-\bar{v}_pt)\right) \label{H}
\end{equation}
where $\bar{v}_p=v_p-v_g$. This Hamiltonian corresponds to a non-integrable, non-autonomous, one-degree of freedom system that describes the complex particle motion. In the rest of this work we drop bars for simplicity of notation. Although the details of the wave-particle interaction may depend on the exact profile of the wavepacket, for the purpose of studying the dependence of particle momentum variation on the main characteristics of a wavepacket, namely its amplitude, spatial width and phase velocity, we consider a Gaussian profile of the form
\begin{equation}
    \Phi(z)=A\exp\left(-\frac{z^2}{2 \sigma^2}\right) \label{Phi}
\end{equation}
where $A$ and $\sigma$ are the wavepacket amplitude and width, respectively. In the following, we will systematically investigate the role of the wavepacket characteristics on the single and the collective particle momentum variation for the entire range of the $(A,\sigma)$ parameter space. Particle dynamics cannot be treated analytically in general, due to the non-integrability of the underlying system. However, there exist domains of the $(A,\sigma)$ parameter space where valid analytical results can be obtained. One of the basic aims of this work is to provide the range of validity of such analytical results under a comparison with results obtained by numerically integrating the Hamiltonian equations of the motion.

\subsection{Small-amplitude wavepackets: near-integrable dynamics}

For a small-amplitude wavepacket the potential term of the Hamiltonian (\ref{H}) can be considered as a perturbation to the kinetic energy term describing the free particle motion, which is integrable, with the energy (and the momentum) of the particle being constants of the motion. The introduction of the potential term renders the system non-integrable due to its explicit time-dependence, and does not allow the existence of an invariant of the motion. However, under small-amplitude perturbations, approximate invariants of the motion can be analytically constructed with the utilization of the Canonical Perturbation Theory (CPT) \cite{Lichtenberg_book}. A near-identity, first-order approximation of a mixed-variable generating function
\begin{equation}
    S(P,z)=Pz+S_1(P,z)
\end{equation}
with 
\begin{equation}
    \frac{\partial S_1}{\partial t}+v_p\frac{\partial S_1}{\partial z}=-\Phi(z)\sin\left(k(z-v_p t)\right)
\end{equation}
provides a first-order approximate invariant of the near-integrable system, as
\begin{equation}
    P=p-\frac{\partial S_1}{\partial z} \label{P}
\end{equation}
For the Gaussian profile (\ref{Phi}), $S_1$ is obtained as

\begin{equation}
    \fl S_1 =-\frac{1}{2i}\sqrt{\frac{\pi}{2}}\frac{A\sigma}{P}e^{-\frac{\sigma^2k^2}{2}\left(1-v_p/P\right)^2}e^{i\frac{kv_p}{P}(z-Pt)}\bigg[1+Erf\bigg(\frac{z-i\sigma^2k\left(1-v_p/P\right)}{\sqrt{2}\sigma}\bigg)\bigg]+c.c. \label{S1}
\end{equation}

The form of the $S_1$ contains information for both single and collective particle dynamics under the presence of a small-amplitude wavepacket of arbitrary width. \cite{Kominis_06} The magnitude of $S_1$ corresponds to the effective perturbation strength of the particle motion under the presence of the wavepacket and depends on all the wavepacket characteristics (amplitude, width and phase velocity) as well as the initial momentum of the particle. Its dependence on the initial particle momentum is particularly strong so that significant momentum variation takes place in the particle momentum space in regions centered around the resonance $p \simeq P=v_p$, with an extent that is inversely proportional to the spatial width ($\sigma$) of the wavepacket. Moreover, the overall strength of the perturbation depends on the product of the potential amplitude ($A$) and the time spent by the particle within the wavepacket ($\sigma/P$).

By setting $t=2\pi/\omega$ $(\omega=kv_p)$ and $P=p$ (valid to first-order) in the rhs of (\ref{P}), the contours of the function $P(p,z)$ provide the single particle orbits in the $(p,z)$ plane as seen in a stroboscopically obtained Poincare surface of section \cite{Lichtenberg_book}. Moreover, the collective particle behavior can be studied on the same basis, by considering an ensemble of charged particles initially located far on the left $(z<0)$ of the wavepacket at a distance equal to several lengths of its spatial width $(\sigma)$ and positions uniformly distributed along an interval equal to several times the wavelength of the carrier wave $(2\pi/k)$, as they collide and pass through the wavepacket. As the wavepacket-particle interaction has a finite duration and is spatially localized, particles entering the interaction region with a given initial momentum and various initial positions, may exit the wavepacket with a varied final momentum. The extreme values of particle momentum variation with respect to its initial position, for a given initial momentum can be obtained from (\ref{P}) by taking the extreme values of $S_1$ with respect to $z$ as
\begin{equation}
    \Delta p_{M,m}=\pm\sqrt{2\pi}A\sigma\frac{kv_p}{p^2}e^{-\frac{\sigma^2k^2}{2}\left(1-v_p/p\right)^2} \label{P_Mm}
\end{equation}
with the plus and minus signs corresponding to maximum (M) and minimum (m) values, respectively.
Moreover, the mean momentum variation and the standard deviation with respect to all initial positions, for a given initial momentum  can be calculated with the utilization of $S_1$ \cite{Kominis_12} with the former given as
\begin{equation}
    \fl <\Delta p>_{z_0}=\frac{1}{2}\frac{\partial}{\partial p} \left< \left(\frac{\partial S_1}{\partial p}\right)^2\right>_z=-\pi A^2\sigma^2\frac{k^2 v_p^2}{p^5}e^{-\frac{\sigma^2k^2}{2}\left(1-v_p/p\right)^2}\left[\sigma^2 k^2 \frac{v_p}{p}\left(1-\frac{v_p}{p}\right) + 2\right] \label{mean}
\end{equation}

Finally, the first-order generating function $S_1$ can be used for providing analytically the modification of an initial momentum distribution function of an ensemble of particles, under interaction with the wavepacket \cite{Kominis_10}.

\subsection{Finite-amplitude, infinite-width wavepackets: integrable dynamics}
For the limiting case of an infinite-width wavepacket, the spatial profile of $\Phi$ reduces to a constant $\Phi=A$ (for the Gaussian profile (\ref{Phi}) this limit corresponds to $\sigma \rightarrow \infty$). In this case, a transformation to the phase velocity frame, renders the system autonomous and integrable. The original Hamiltonian is written as
\begin{equation}
    H(p,z)=\frac{p^2}{2}+A\sin(kz) \label{ONeil}
\end{equation}
resembling the Hamiltonian of a pendulum, and the equations of motion can be solved in terms of Jacobi elliptic functions \cite{ONeil}, providing analytical results for single as well as collective particle dynamics, for arbitrary wave amplitude $A$. 

Despite the essential differences between the dynamics of the original system and (\ref{ONeil}), a qualitative correspondence between the two systems can be based on the relation between the maximum momentum variations for a finite-width wavepacket with an appropriate $\sigma$, as estimated from the first-order perturbative result (\ref{P_Mm}), and an infinite-width wavepacket with the same amplitude ($A$), as measured at the separatrix of the system (\ref{ONeil}), $\Delta p _{M,m}=\pm2\sqrt{A}$. By equating these momentum variations, we obtain the appropriate width $\sigma$ as
\begin{equation}
    \sigma=\frac{v_p}{k}\sqrt{\frac{2}{\pi A}} \label{sigma_ON}
\end{equation}

\section{Numerical Results and Discussion}
As an inherently nonlinear process, the interaction of particles with a wavepacket of a finite spatial width, depends strongly on the effective interaction strength which is proportional to both the amplitude $A$ and the width $\sigma$ of the wavepacket. For a relatively small effective interaction strength the analytical results of the Canonical Perturbation Theory are expected to be in qualitative and quantitative agreement with those obtained from direct numerical calculations, whereas for stronger interactions quantitative and, eventually, qualitative differences are expected. In the following, we investigate single and collective particle momentum variations for ensembles of particles passing through the wavepacket via numerical integration of the underlying equations of motion, by utilizing a standard 4-th order Runge-Kutta method, and we obtain the domains of validity of the analytical results. Moreover, we illustrate the emergence of an increasingly complex dynamical behavior as the effective interaction strength increases, giving rise to strongly nonlinear interactions. It is worth emphasizing that the one-degree-of-freedom character of the system facilitates the systematic investigation of the collective particle behavior due to the fact that the two-dimensional phase space can be densely populated with initial conditions without a significant computational cost, as in the case of more degrees of freedom. 

We consider an ensemble of particles with their initial positions uniformly distributed along a region spanning several wavelengths of the carrier wave. The ensemble is located far on the left of the wavepacket $(z<0)$ at a distance much larger than its spatial width, and has initial momenta $p_0$.
The wavepacket is considered to have a wavenumber $k=1$ and phase velocity $v_p=1$, which is equivalent to normalizing space to the wavenumber and measuring velocities (and momenta) in terms of the phase velocity of the wavepacket, whereas time is normalized to the frequency of the wave. \cite{Fuchs_85, Menyuk_85, Bruhwiler_92}        

The momentum variations of these particles are shown in Fig. 1 and Fig. 2 for  small ($A=10^{-4}$) and larger ($A=10^{-2}$) amplitude of the wavepacket, respectively, as obtained by numerically solving the underlying equations of motion. The momentum variation of the particles, after their transition through the wavepacket, is shown as a function of their initial momentum $p_0$ for various spatial widths $(\sigma)$ and $v_p=1$ (note that this is the phase velocity in the group velocity frame of reference). In all cases, significant momentum variation takes place for initial momenta in the vicinity of the phase velocity, in accordance to the resonant character of the interaction. The width of the initial momentum interval for significant momentum variation is inversely proportional to the spatial width of the wavepacket; wider wavepackets have a narrower spectral content resulting in sharper resonances with the particles. For both small (Fig. 1) and larger (Fig. 2) amplitudes of the wavepacket, the effective interaction strength increases with $\sigma$ and the scatter plots of the momentum variations change symmetry and demonstrate a stronger non-uniformity. Particles with the same initial momentum $p_0$ but different initial position undergo different momentum variations due to different relative phases with the wave variations. The extreme and mean momentum variations as well as the standard deviations of distributions of particles with different initial positions and initial momentum $p_0$ are depicted in Figs. 3, 4 and 5, respectively. A remarkable agreement between numerical and analytical results, obtained via the canonical perturbation approach [equations (\ref{P_Mm}), (\ref{mean})] , is shown for relatively small effective interaction strengths ($\sim A\sigma$), as depicted in Figs. 3(a), 4(a), 5(a), as well as for smaller interaction strengths (not shown here). For larger perturbation strengths [Figs. 3(b-c), 4(b-c), 5(b-c)], there are significant quantitative differences between numerical and analytical results. The extreme values of the momentum variations shown in Fig. 3 present a different dependence on the initial momenta $p_0$ as the effective perturbation strength increases, with the maximum variations appearing for initial momenta at the limits of the region centered around $p_0=v_p=1$ where significant variation takes place. An increasing complexity of the dependence of the mean values and the standard deviations on the initial momentum is also shown for large interaction strengths [Figs. 4(d-f), 5(d-f)] corresponding to the particle "bunching" shown in the scatter plots [Figs. 1(c-e), 2(c-e)]. These effects manifest the eventual transition from regular to chaotic particle dynamics related to stronger dependence on the initial conditions and non-smooth particle distribution functions, due to an increasing perturbation strength.   

Analytical results for the mean momentum variation as obtained from the analytical solutions of the pendulum-like system (\ref{ONeil}) are compared with numerical results in Fig. 6(a), showing the correspondence between the interaction with a wavepacket of an infinite wave and a finite-width wavepacket with the appropriate $(\sigma)$ given by the condition (\ref{sigma_ON}). The domain of validity of the results obtained from the first-order canonical perturbation method in the $(A,\sigma)$ parameter space is clearly shown in Fig. 6(b), as obtained within a limit of maximum ten percent relative difference with respect to numerical results for the maximum mean momentum variation. The condition (\ref{sigma_ON}) is fulfilled close to the boundary of the domain of validity. The validity of the analytical results is restricted to smaller values of the amplitude $(A)$ as the spatial width $(\sigma)$ increases, whereas for small spatial widths the analytical results remain valid even for relatively large values of the amplitude. It is worth mentioning, that this domain of validity of the first-order perturbation theory refers to the collective particle dynamics (mean momentum variations). However, the effective perturbation strength, as obtained from the magnitude of $S_1$ in Eq. (\ref{S1}), depends strongly on the initial momentum of the particle - namely the same wavepacket acts on different particles of an ensemble with different strength; therefore, the validity of the analytical perturbation theory may have different limits for different single particles. The domain of validity of the analytical results of the Canonical Perturbation Theory, can be further extended in the parameter space to regions of larger effective perturbation strength by utilizing higher-order calculations \cite{Lichtenberg_book} capturing the aforementioned qualitative differences between the first-order analytical and numerical results depicted in Figs. 3-5 and enabling a better quantitative agreement. The investigation of the validity of such analytical results in this one-degree-of-freedom system provides intuition and guidelines for their validity in higher-degree-of-freedom systems describing particle dynamics under the presence of homogeneous or inhomogeneous magnetic fields as well as electromagnetic waves, where the respective numerical investigations have a significant computational cost.

A significant qualitative difference between analytical and numerical results is related to the mean momentum variation of an ensemble of particles with uniformly distributed positions and momenta under their interaction with a localized wavepacket. According to the analytical results, obtained via the Canonical Perturbation Theory and valid for small-amplitude wavepackets, the mean momentum variation (averaged over both initial positions and momenta) is equal to zero. This is equivalent with the condition that the algebraic value of the overall area between the analytically obtained curves in Fig. 4(a-c), corresponding to the integration over all initial momenta, is equal to zero. Moreover, this is in accordance to the standard Landau resonance results \cite{Dawson_61, Drummond_04} where such a momentum exchange between the wave and particle is zero when the slope of the momentum distribution function is zero at the resonant particle momentum value. In Figs. 7(a,b), the numerically calculated total average particle momentum and energy, for a uniform initial position and momentum distribution function, are shown along with their dependence on the wavepacket width $(\sigma)$. It is clear that there is a significant average momentum and energy variation, that varies with $\sigma$ and forms a plateau for sufficiently large $\sigma$. This is a purely nonlinear effect which can be considered as a nonlinear Landau resonance mechanism \cite{ONeil, Brodin_97, Manfredi_97, Pegoraro_00, Wang_13}. The calculation of the mean momentum gain and loss, depicted in Figs. 7(c,d) clearly shows the asymmetry of the two curves resulting in an excess momentum gain. The symmetric analytical results, along with the value of $\sigma$ corresponding to the curve (\ref{sigma_ON}) in the $(A, \sigma)$ parameter space in Fig. 6(b), are also shown. The nonlinear character of this effect cannot be captured even for relatively small-amplitude wavepackets by a first-order perturbation theory, and necessitates higher-order calculations.

In all previous analytical and numerical calculations, uniform initial momentum distributions have been considered, in order to investigate the role of the initial momentum on the momentum variation due to particle transition through a localized wavepacket. In addition to this purpose, results for a uniform initial momentum distribution can be utilized to study the modification of any nonuniform initial momentum distribution under interaction with a localized wavepacket, by considering it as a sum of "beamlets" with different initial momentum and uniformly distributed initial positions, and assigning a specific "weight" corresponding to a different number of particles (macro-particles) according to the nonuniform distribution function of interest. In such way, based on a given set of calculations for a uniform momentum distribution, results for any momentum distribution can be obtained. The modification of a Maxwellian initial momentum distribution $f(p) \sim \exp\left[-(p-p_0)^2/2\sigma_{th}^2\right]$, with $\sigma_{th}^2=0.065$ and $p_0=1$, under interaction with a wavepacket with $(A,\sigma)=(0.015,15)$ and $v_p=1.35$ is depicted in Fig. 8, as obtained from a weighted uniform and a random Maxwellian initial momentum distribution.

The effect of the interaction of a localized wavepacket with an ensemble of particles with nonuniformly distributed initial momenta, is illustrated in Fig. 9, for a Maxwellian initial distribution. The time evolution of the initially Maxwellian distribution is depicted in Fig. 9(a). The modification of the initial distribution function takes place on a finite time interval during particles' passage through the wavepacket resulting in a final momentum distribution after their exit. The dependence of the final momentum distribution on the phase velocity (in the group velocity frame) $v_p$ of the wave is shown in Fig. 9(b), where strong modification is shown for particle momenta close to $v_p$. Finally, the dependence of the final momentum distribution on the width $(\sigma)$ and the amplitude $A$ of the wavepacket, is depicted in Figs. 9(c) and (d), respectively. The modification increases with the interaction strength (either for increasing amplitude or width) and it saturates for larger $\sigma$ whereas it continuously increases for larger $A$, resulting in a significant secondary maximum of the momentum distribution function.

\section{Summary and Conclusions}
The fundamental interaction of charged particles with electrostatic wavepackets of finite spatial width has been studied in terms of single and collective momentum variations. The parameter space consisting of the wavepacket amplitude, width and group/phase velocity has been investigated in terms of the particle momentum and energy variation of particles passing through the wavepacket. The main findings of this work are the systematic investigation of the range of validity of perturbative analytical results and the identification of the emerging complexity of the collective particle dynamics for stronger perturbations, under an approach that relates single with collective particle dynamics by systematically dissecting an ensemble of particles in terms of initial particle momenta. 

Analytical results, are shown in excellent agreement with numerical calculations, within a range of validity, that is well-defined in terms of an effective interaction strength. The detailed comparison between analytical and numerical results in the one-dimensional model considered in this work, also provides insight and guidelines for the range of validity of the analytical results, in higher-dimensional configurations and/or interactions with electromagnetic waves, where numerical calculations have a significantly higher computational cost. 

Moreover, the increasing complexity of the nonlinear interaction for sufficiently large interaction strength has been identified through the fine structure of the momentum variation dependence on the initial particle momentum. Also, the purely nonlinear mechanism leading to a nonzero momentum and energy transfer to a uniform initial momentum distribution, beyond the prediction of the Landau resonance mechanism, has been described. Finally, the role of the wavepacket characteristics on the modification of the final momentum distribution function of the particles after interacting with the wavepacket has been systematically investigated. The results can also serve as a basis for studying successive particle interactions with the same or different wavepackets under an interative scheme. 

\ack
The authors acknowledge useful discussions on wave-particle interaction dynamics with Dr. Panagiotis Zestanakis and Dr. Giorgos Anastassiou. This work has been carried out within the framework of the EUROfusion Consortium, funded by the European Union via the Euratom Research and Training Programme (Grant Agreement No 101052200 — EUROfusion). Views and opinions expressed are however those of the author(s) only and do not necessarily reflect those of the European Union or the European Commission. Neither the European Union nor the European Commission can be held responsible for them.

\clearpage

\bibliography{bibliography}

\begin{figure*}\centering
\begin{center}
    \subfigure[]
        {\includegraphics[width=\scl\columnwidth]{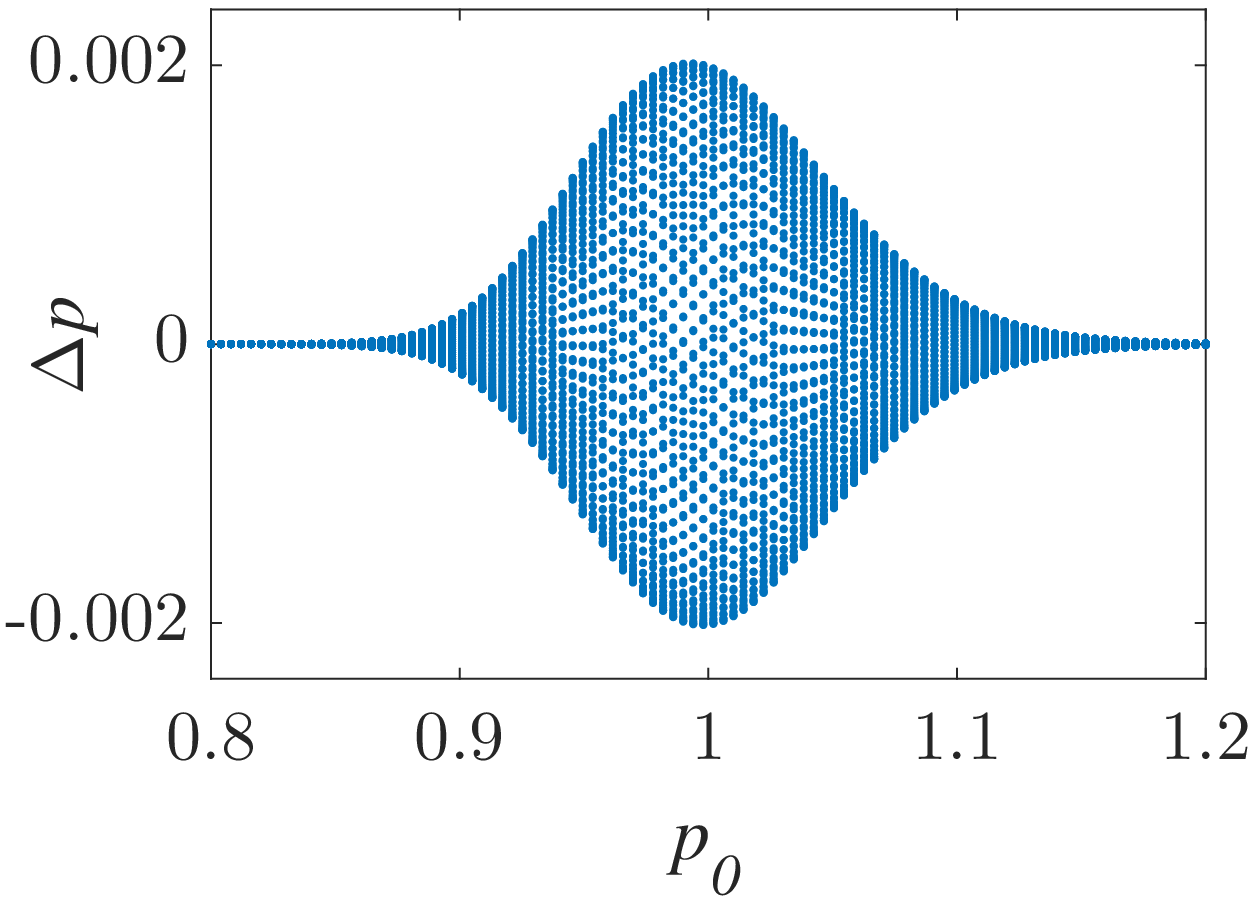}}
    \subfigure[]
        {\includegraphics[width=\scl\columnwidth]{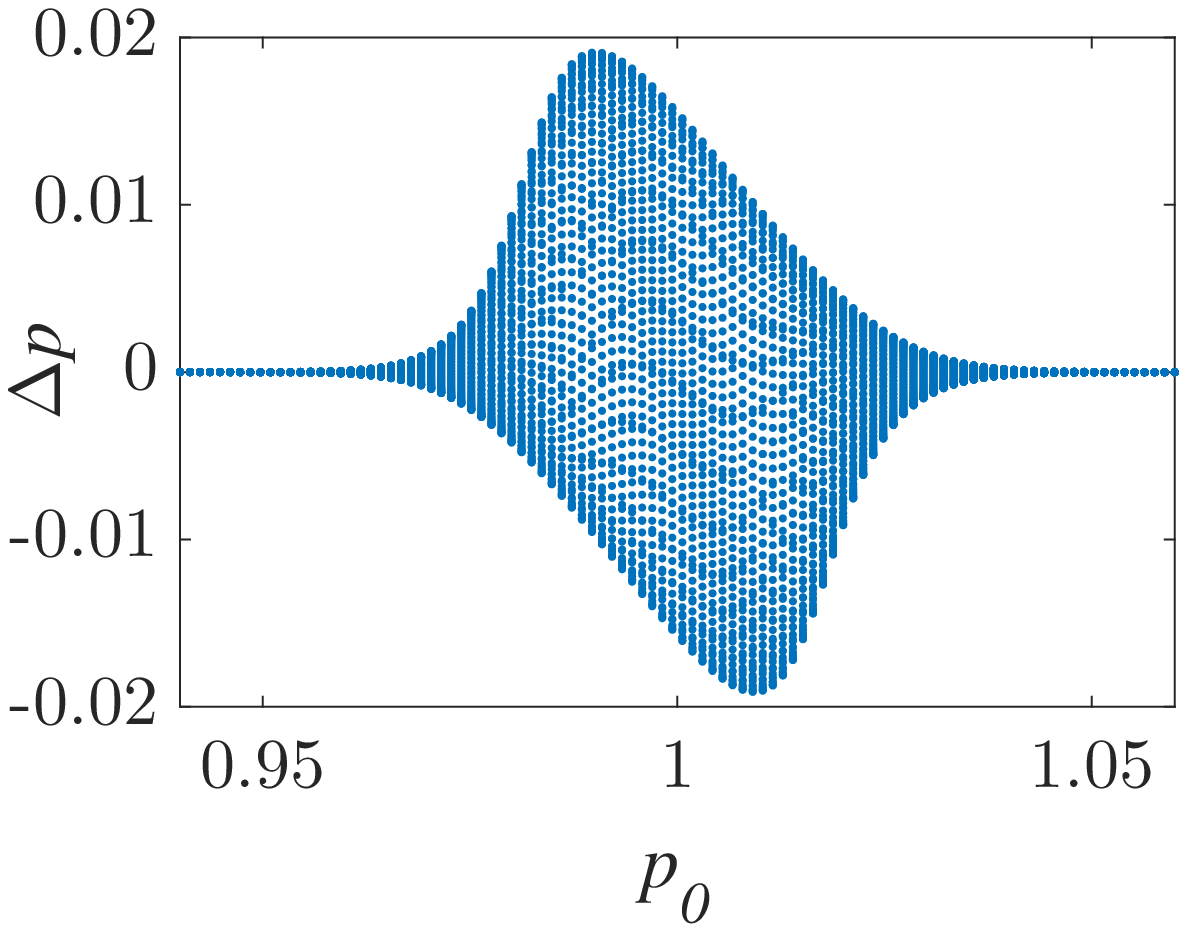}}
    \subfigure[]
        {\includegraphics[width=\scl\columnwidth]{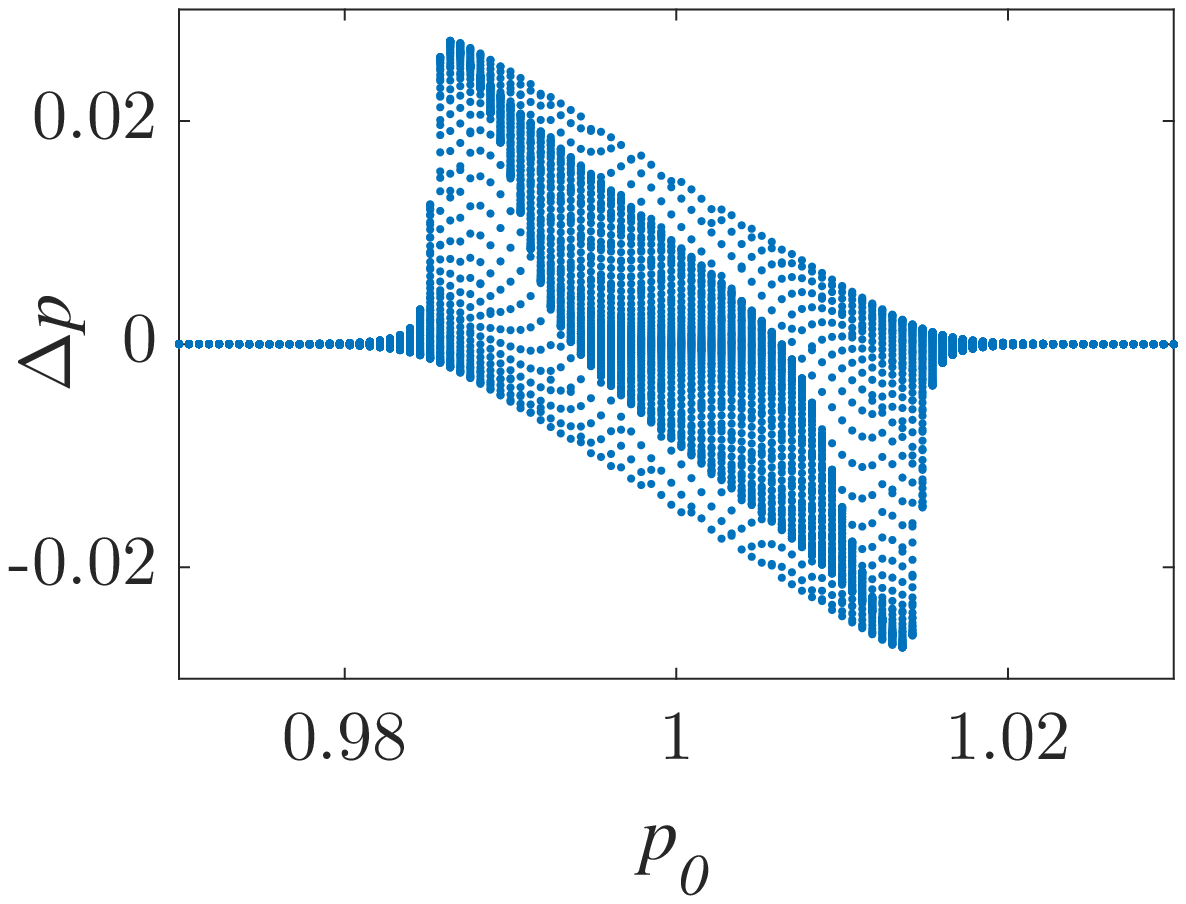}}
    \subfigure[]
        {\includegraphics[width=\scl\columnwidth]{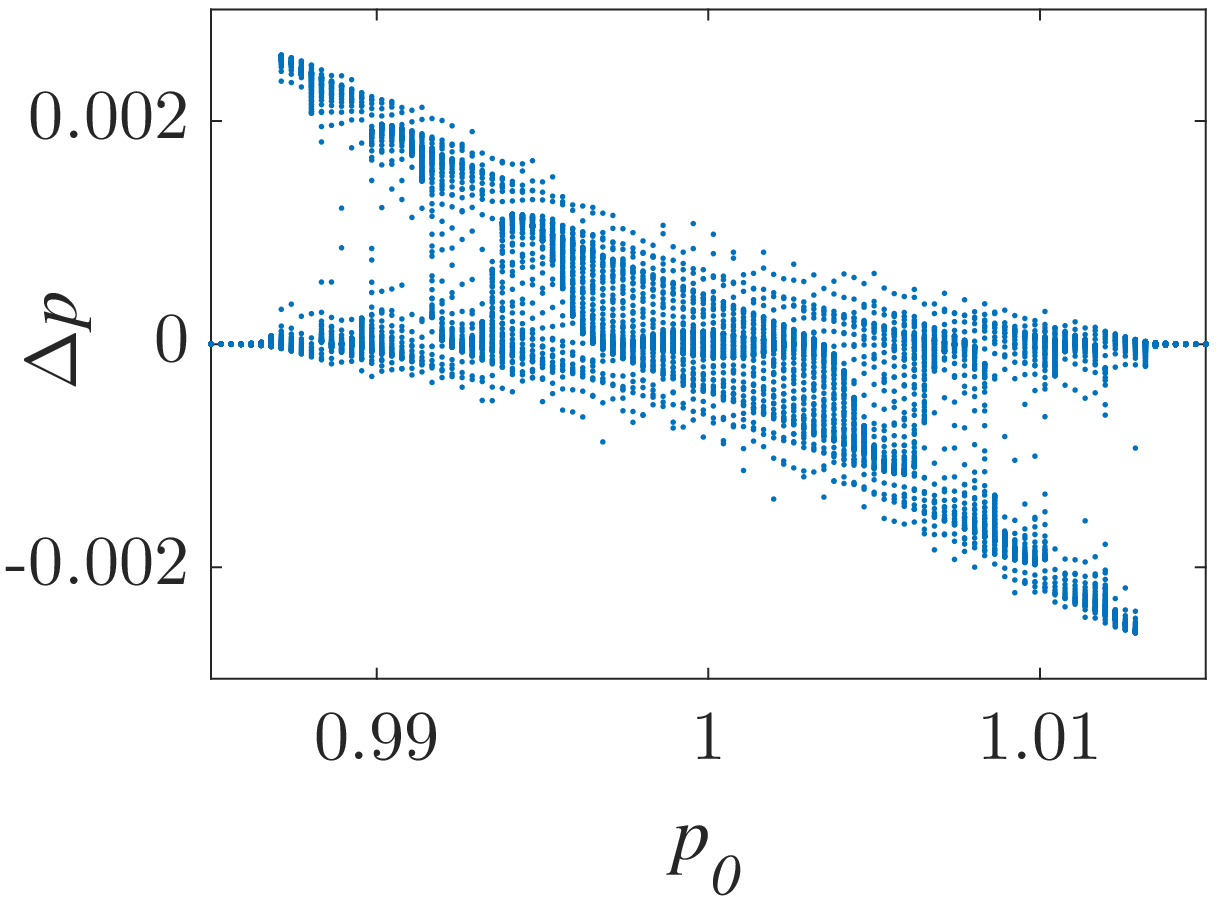}}
    \subfigure[]
        {\includegraphics[width=\scl\columnwidth]{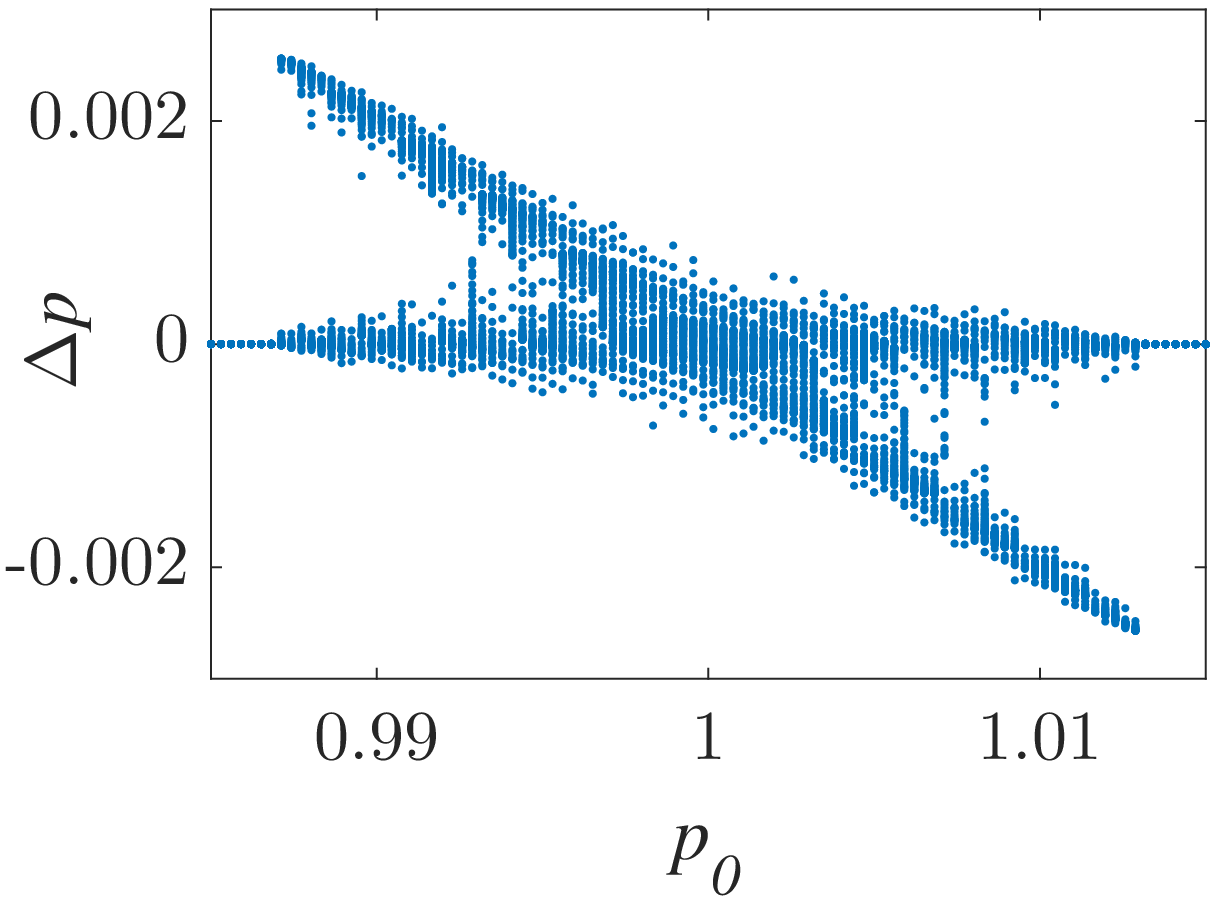}}
    \caption{Momentum variation $\Delta p$ as a function of the initial momentum $p_0$ for an ensemble of particles after a transition through a wavepacket with relatively small amplitude $A=10^{-4}$, $v_p=1$, and various spatial widths $\sigma=20, 80, 200, 600, 1000$ (a-e). The initial positions are uniformly distributed within a region spanning several wavelengths of the carrier wave that is located far on the left of the wavepacket $(z<0)$ at a distance much larger than its spatial width. }
\end{center}
\end{figure*}

\begin{figure*}\centering
\begin{center}
    \subfigure[]
        {\includegraphics[width=\scl\columnwidth]{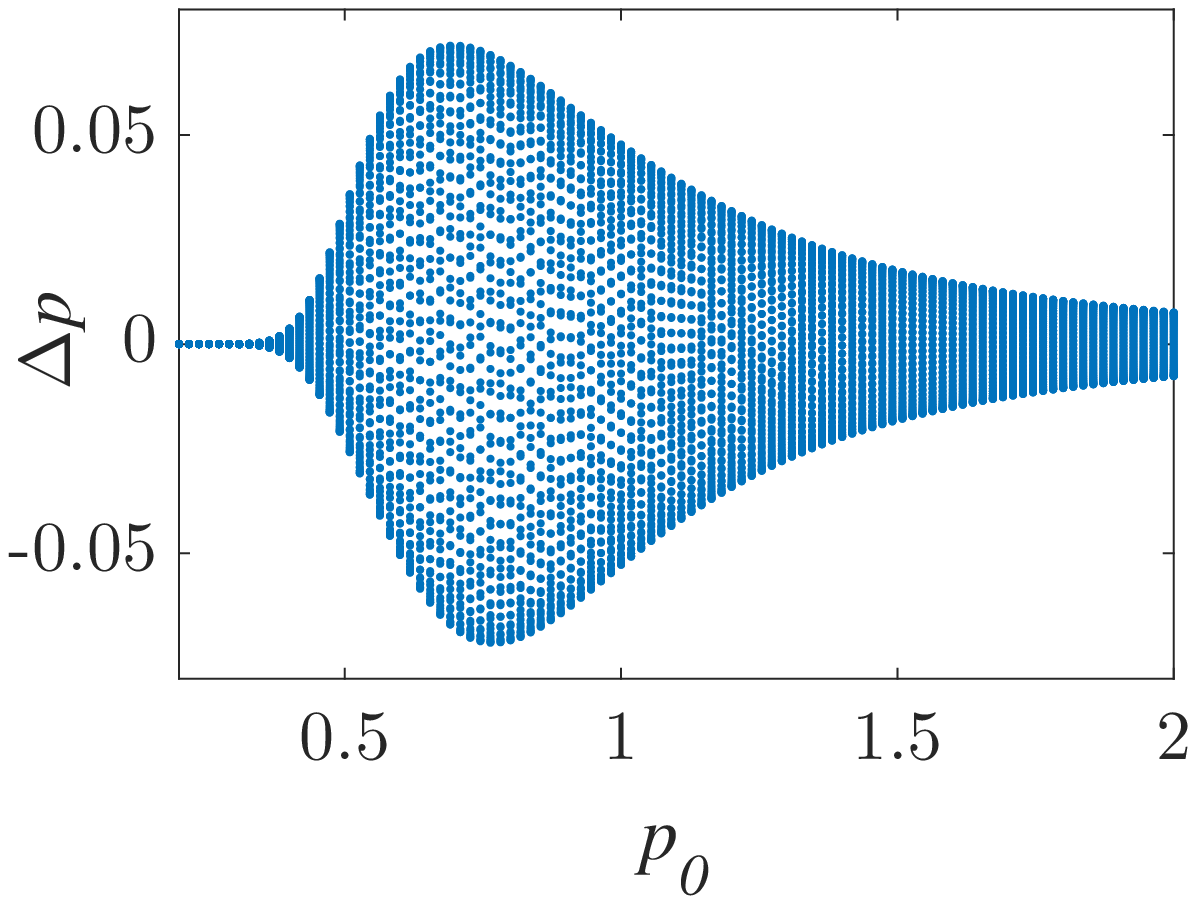}}
    \subfigure[]
        {\includegraphics[width=\scl\columnwidth]{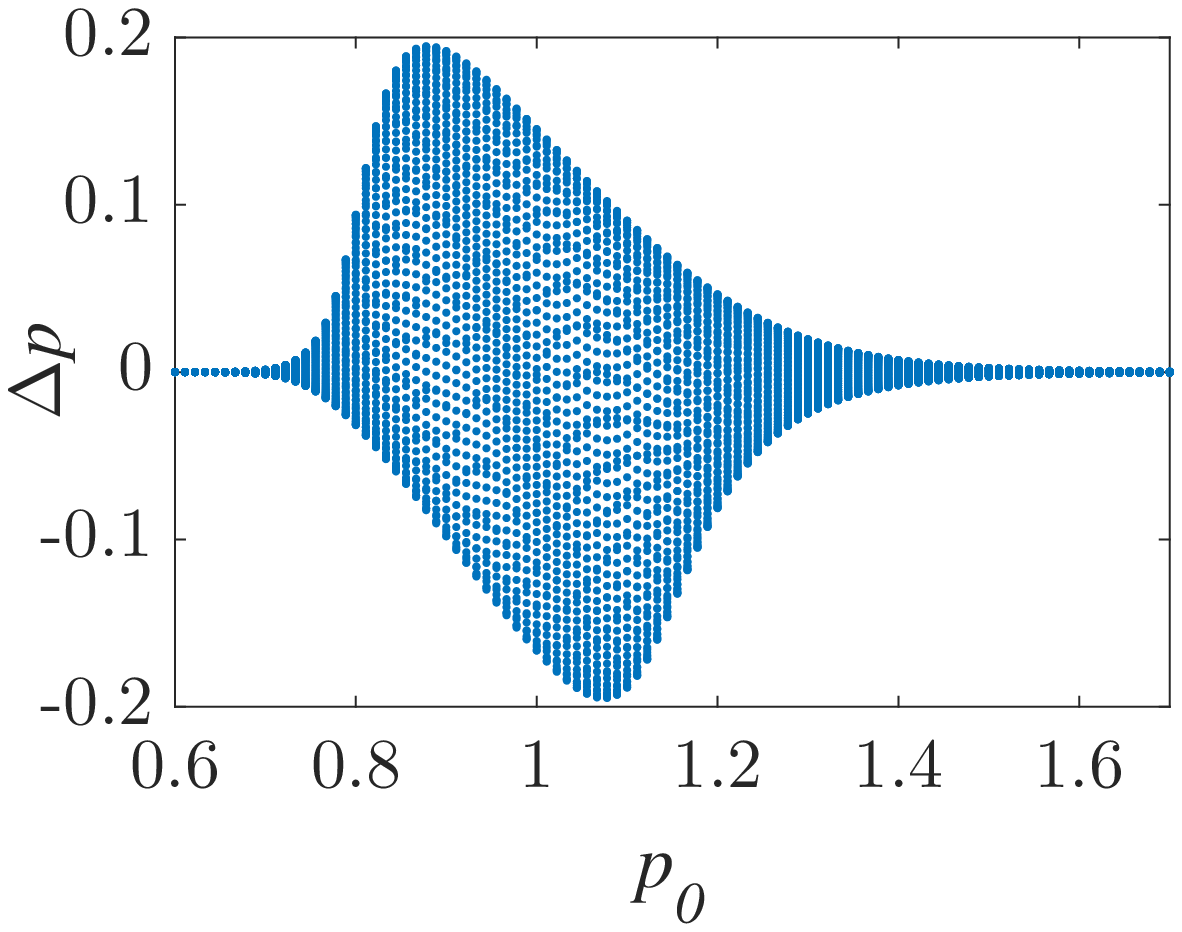}}
    \subfigure[]
        {\includegraphics[width=\scl\columnwidth]{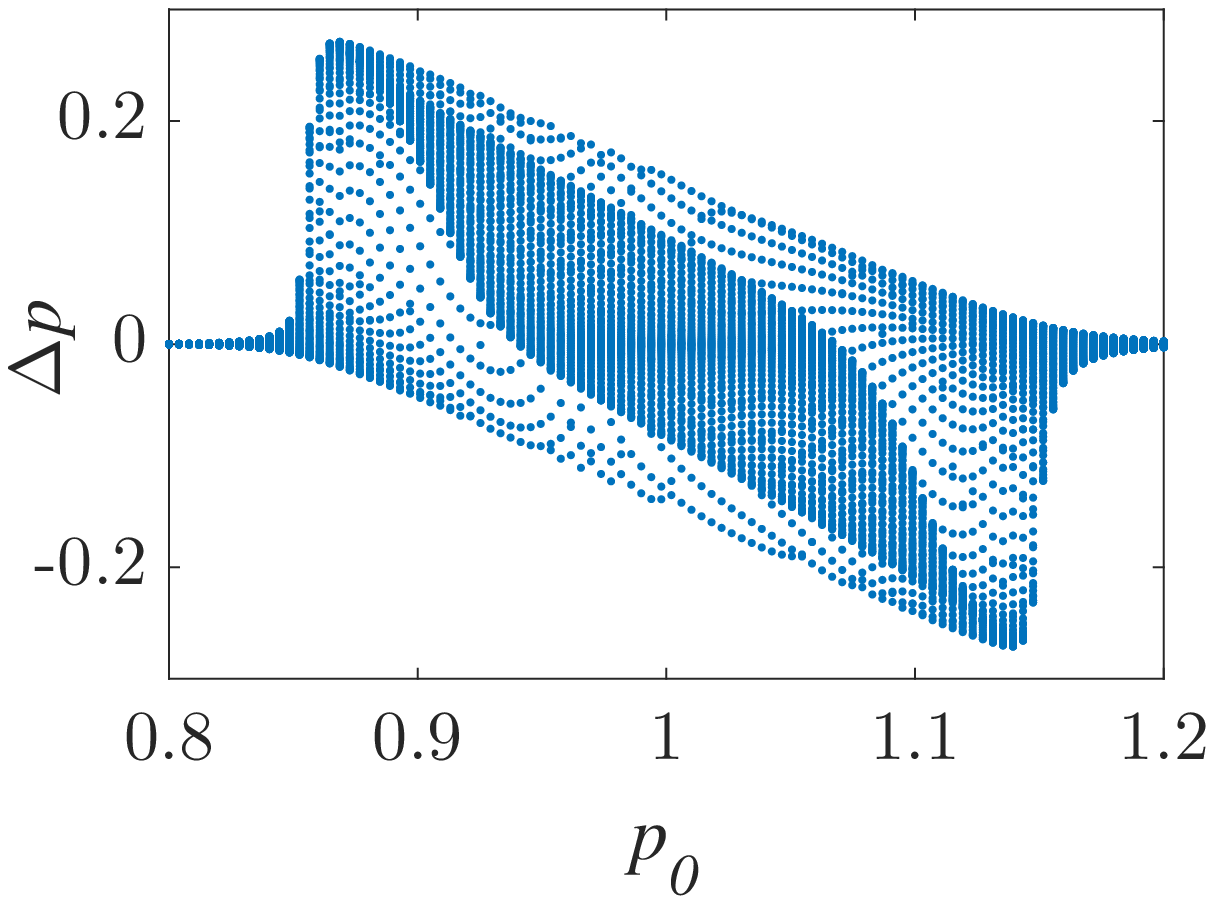}}
    \subfigure[]
        {\includegraphics[width=\scl\columnwidth]{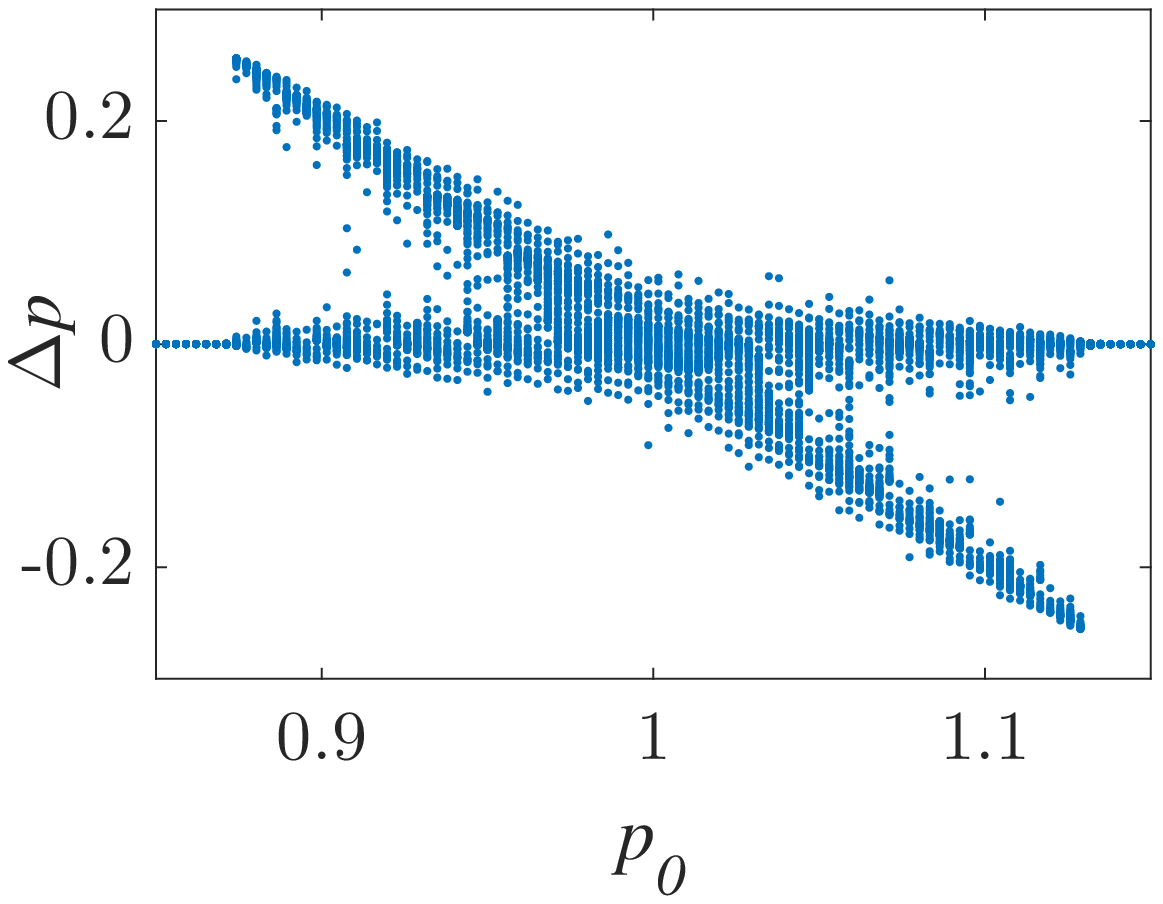}}
    \subfigure[]
        {\includegraphics[width=\scl\columnwidth]{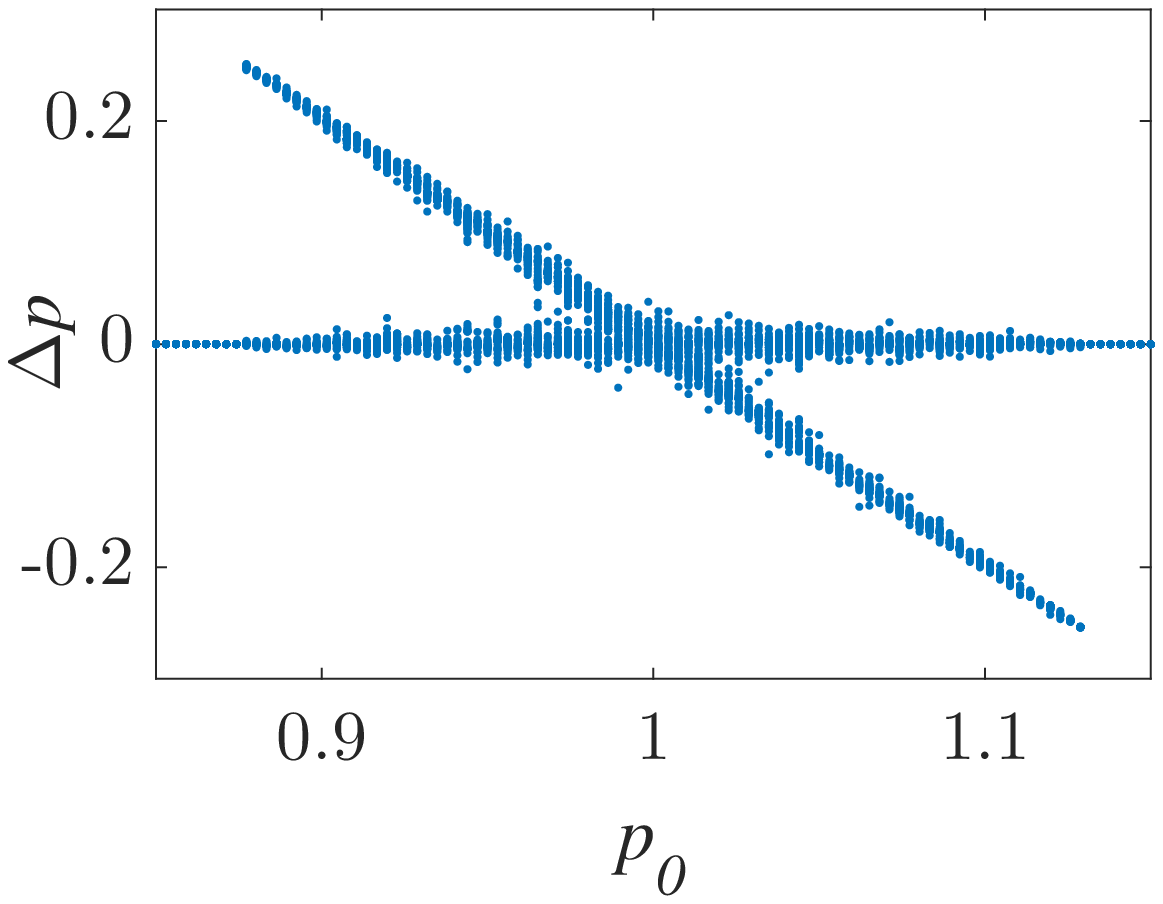}}
        \caption{Momentum variation $\Delta p$ as a function of the initial momentum $p_0$ for an ensemble of particles after a transition through a wavepacket with relatively large amplitude $A=10^{-2}$, $v_p=1$, and various spatial widths $\sigma=2, 8, 20, 100, 300$ (a-e). The initial positions are uniformly distributed within a region spanning several wavelengths of the carrier wave that is located far on the left of the wavepacket $(z<0)$ at a distance much larger than its spatial width.}
\end{center}
\end{figure*}

\begin{figure*}\centering
\begin{center}
    \subfigure[]
        {\includegraphics[width=\scl\columnwidth]{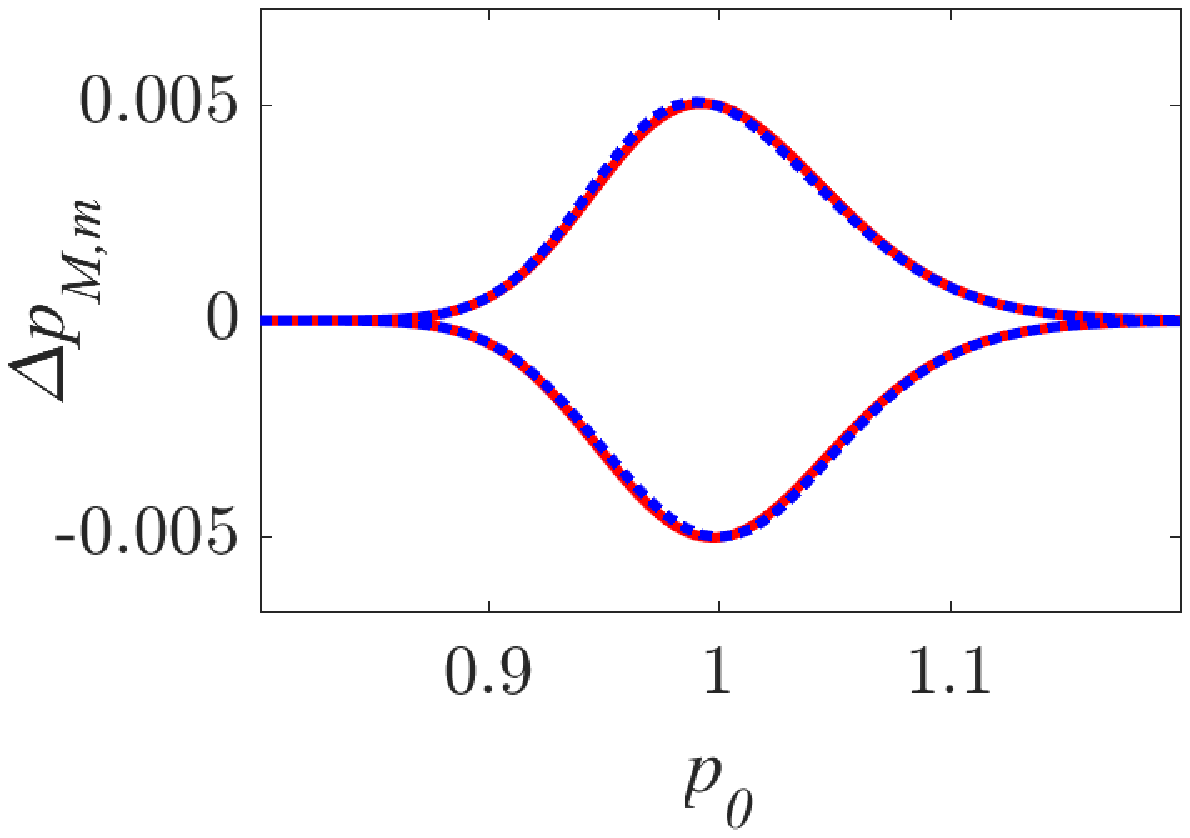}}
    \subfigure[]
        {\includegraphics[width=\scl\columnwidth]{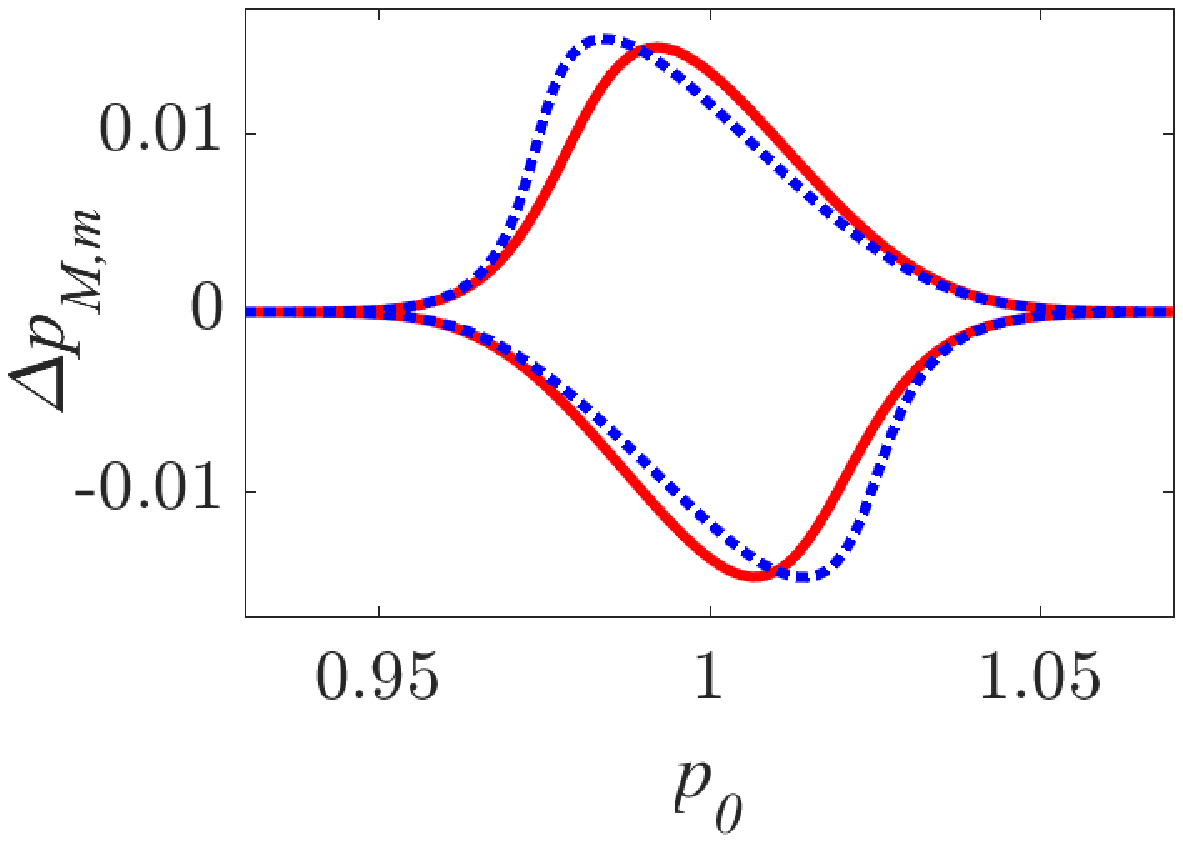}}
    \subfigure[]
        {\includegraphics[width=\scl\columnwidth]{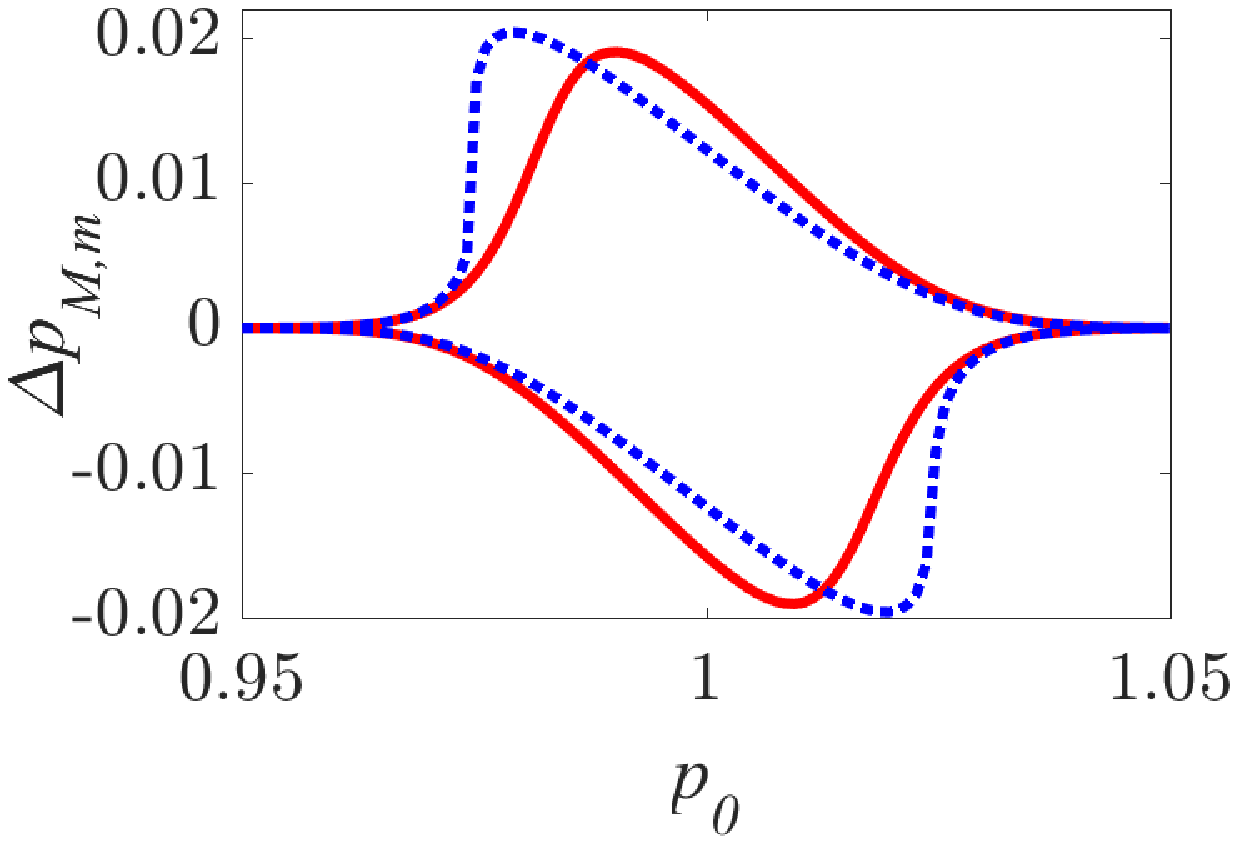}}
    \subfigure[]
        {\includegraphics[width=\scl\columnwidth]{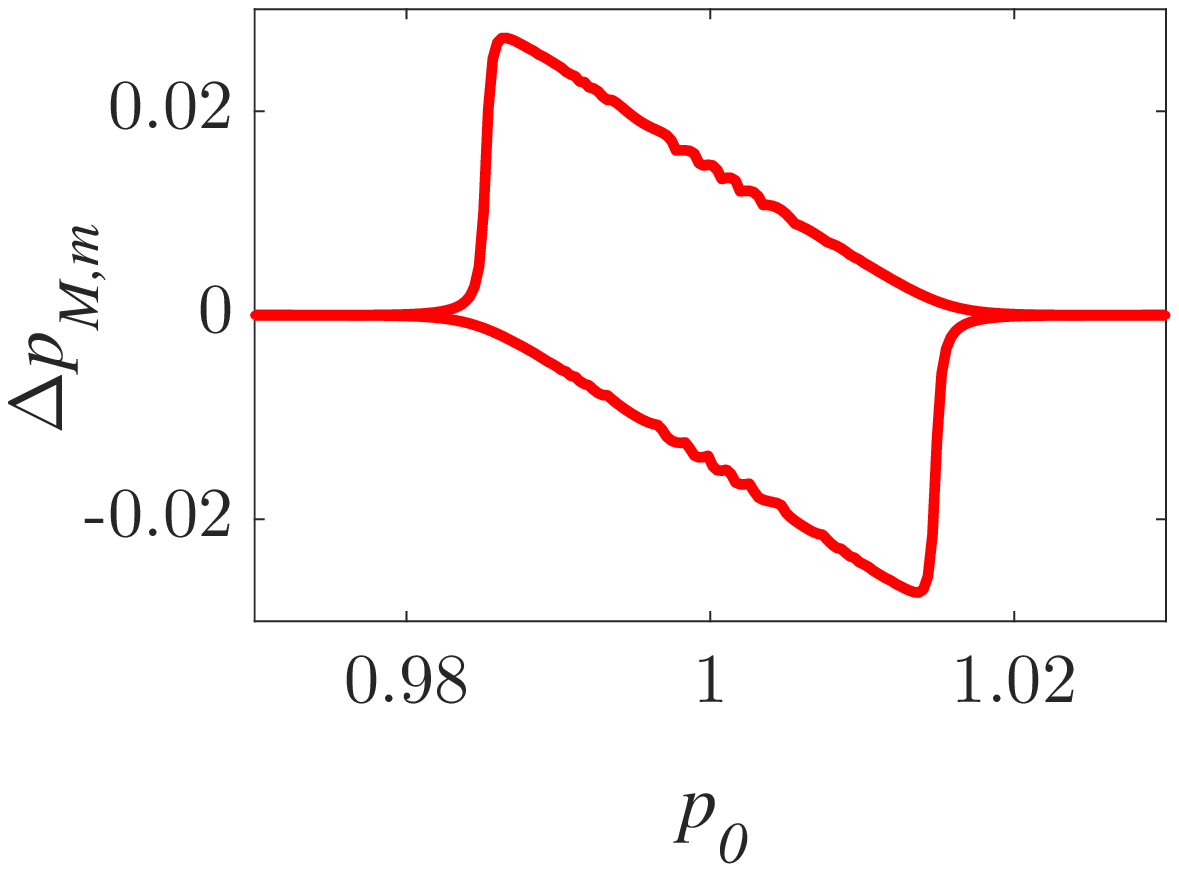}}
    \subfigure[]
        {\includegraphics[width=\scl\columnwidth]{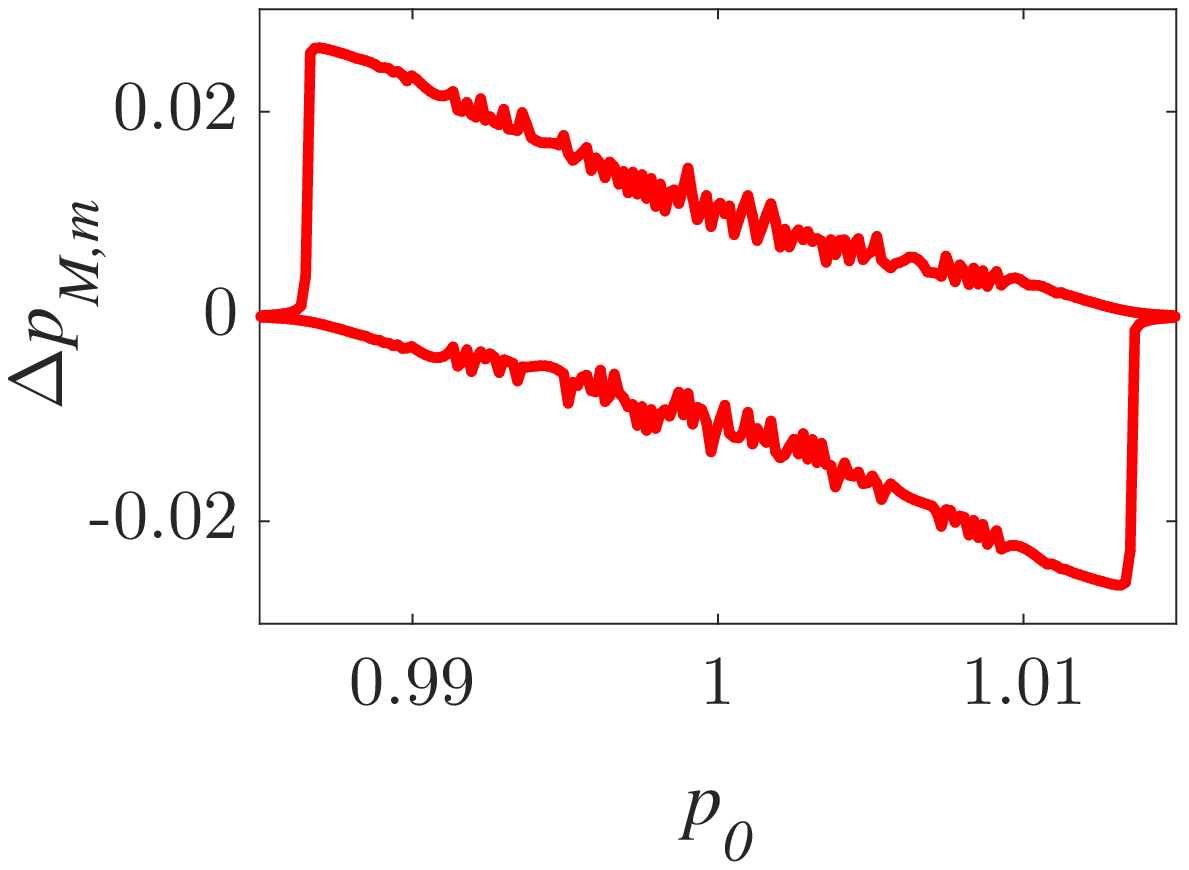}}
    \subfigure[]
        {\includegraphics[width=\scl\columnwidth]{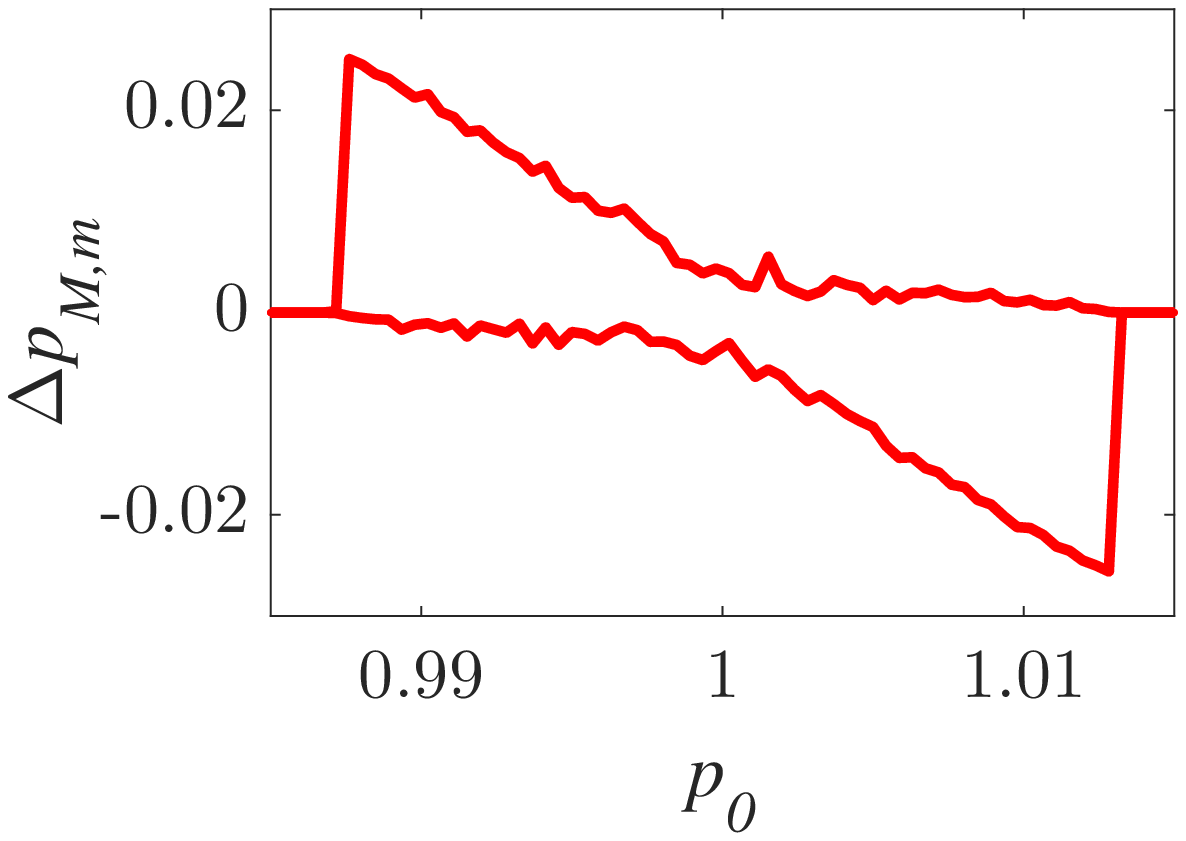}}
    \caption{Numerically (red, solid lines) and analytically (blue, dotted lines) calculated extreme values of momentum variations $\Delta p_{M,m}$ for an ensemble of particles with initial momentum $p_0$ and uniformly distributed initial positions after a transition through a wavepacket with relatively small amplitude $A=10^{-4}$, $v_p=1$, and various spatial widths $\sigma=20, 60, 80, 200, 400, 1500$ (a-f).}
\end{center}
\end{figure*}

\begin{figure*}\centering
\begin{center}
    \subfigure[]
        {\includegraphics[width=\scl\columnwidth]{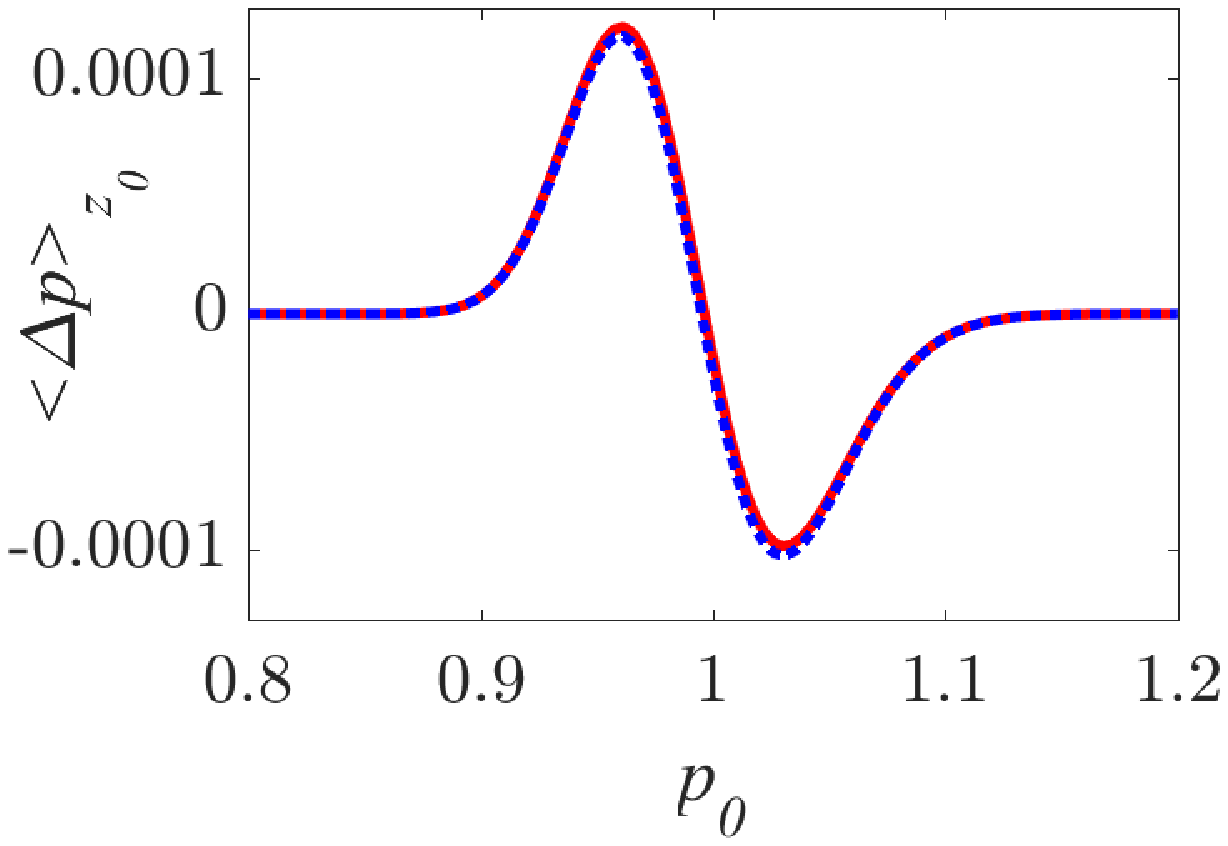}}
    \subfigure[]
        {\includegraphics[width=\scl\columnwidth]{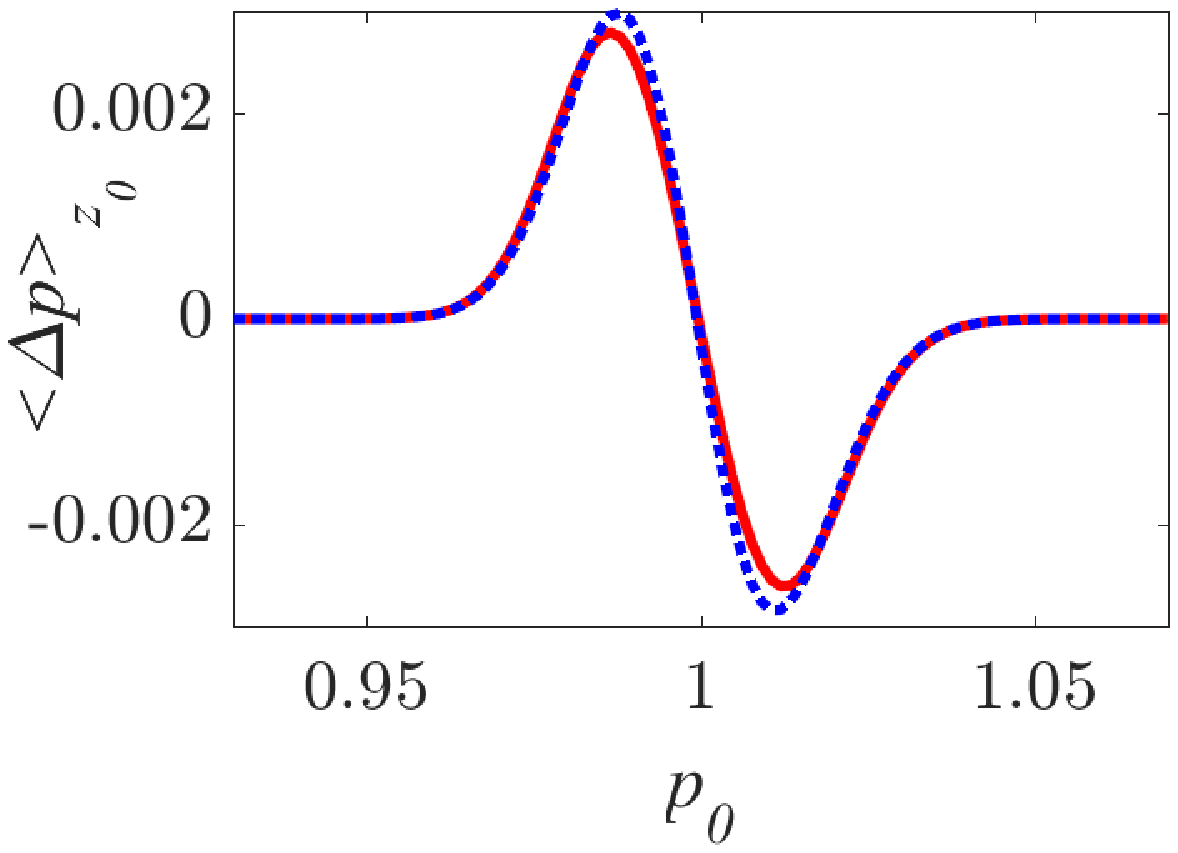}}
    \subfigure[]
        {\includegraphics[width=\scl\columnwidth]{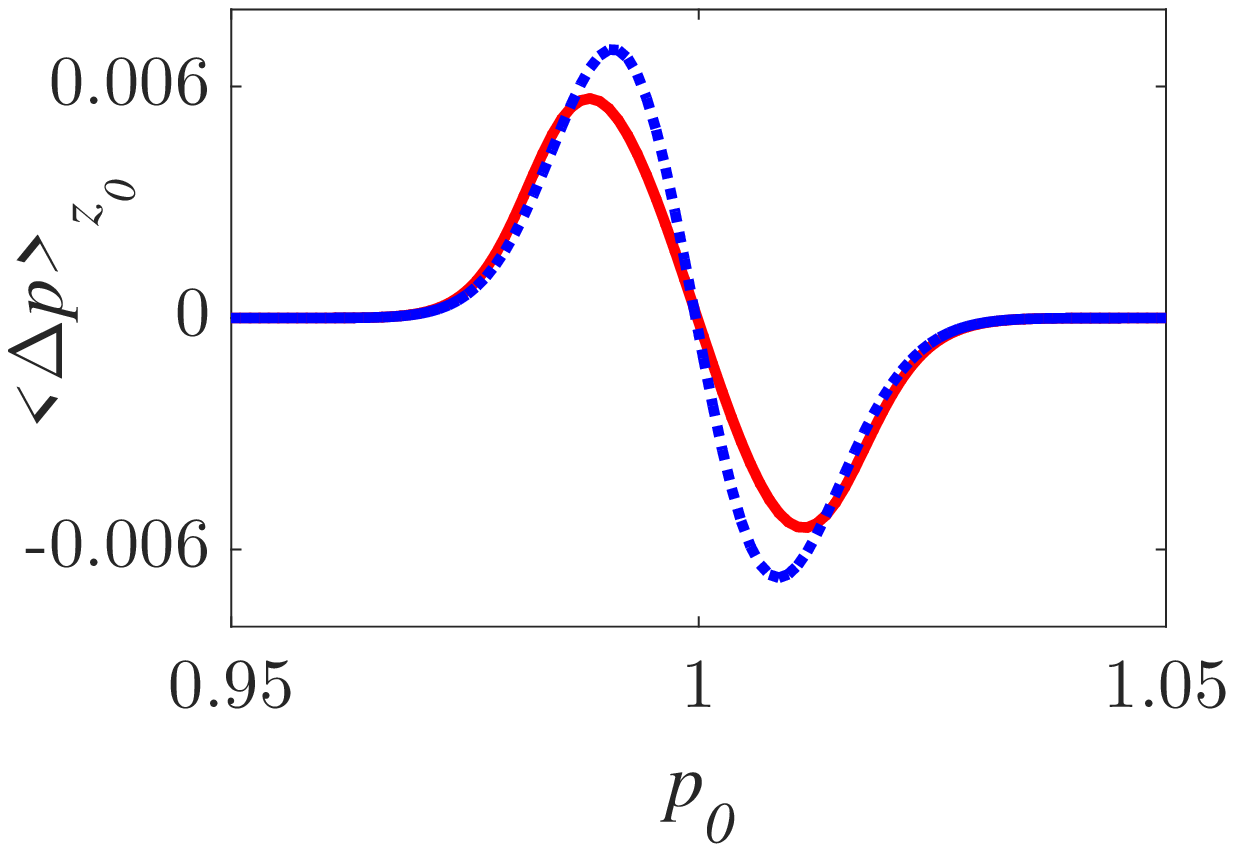}}
    \subfigure[]
        {\includegraphics[width=\scl\columnwidth]{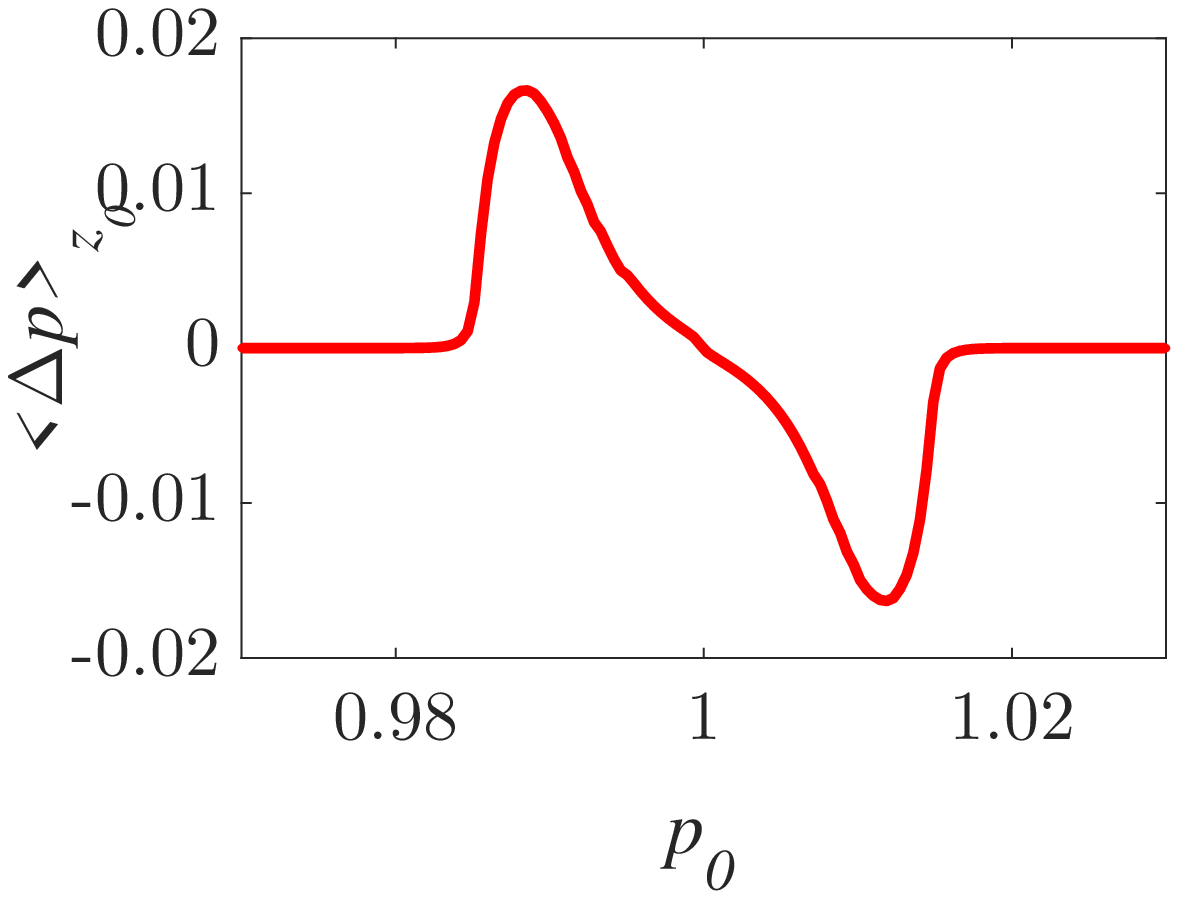}}
    \subfigure[]
        {\includegraphics[width=\scl\columnwidth]{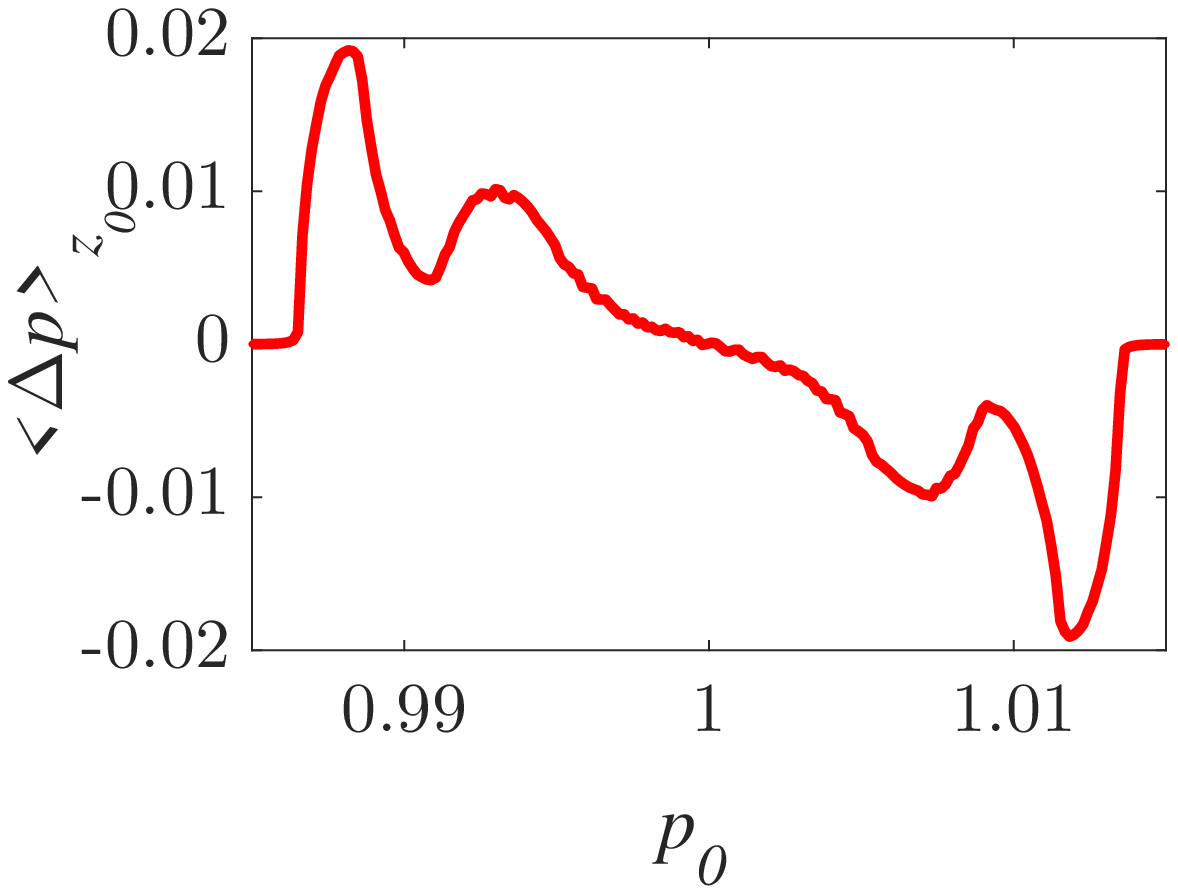}}
    \subfigure[]
      {\includegraphics[width=\scl\columnwidth]{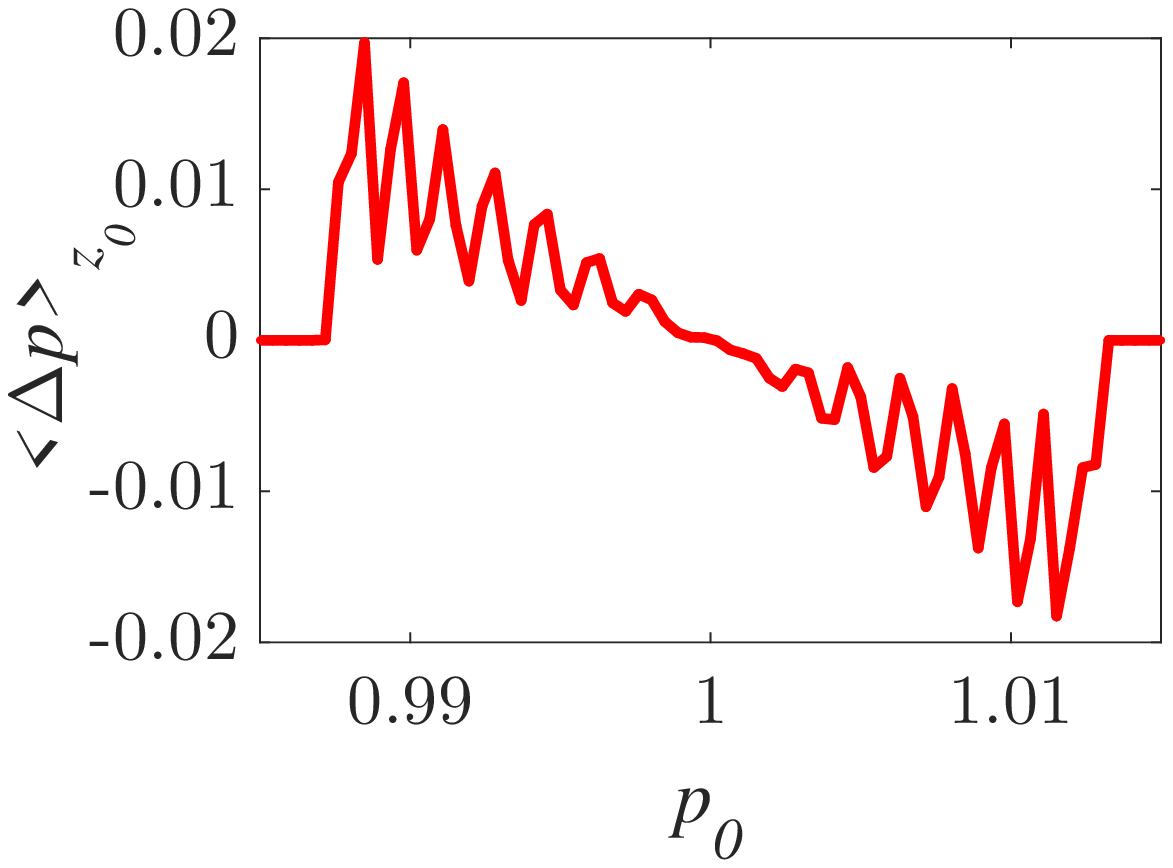}}
    \caption{Numerically (red, solid lines) and analytically (blue, dotted lines) calculated mean values of momentum variations $<\Delta p>_{z_0}$ for an ensemble of particles with initial momentum $p_0$ and uniformly distributed initial positions after a transition through a wavepacket with relatively small amplitude $A=10^{-4}$, $v_p=1$, and various spatial widths $\sigma=20, 60, 80, 200, 400, 1500$ (a-f).}
\end{center}
\end{figure*}

\begin{figure*}\centering
\begin{center}
    \subfigure[]
        {\includegraphics[width=\scl\columnwidth]{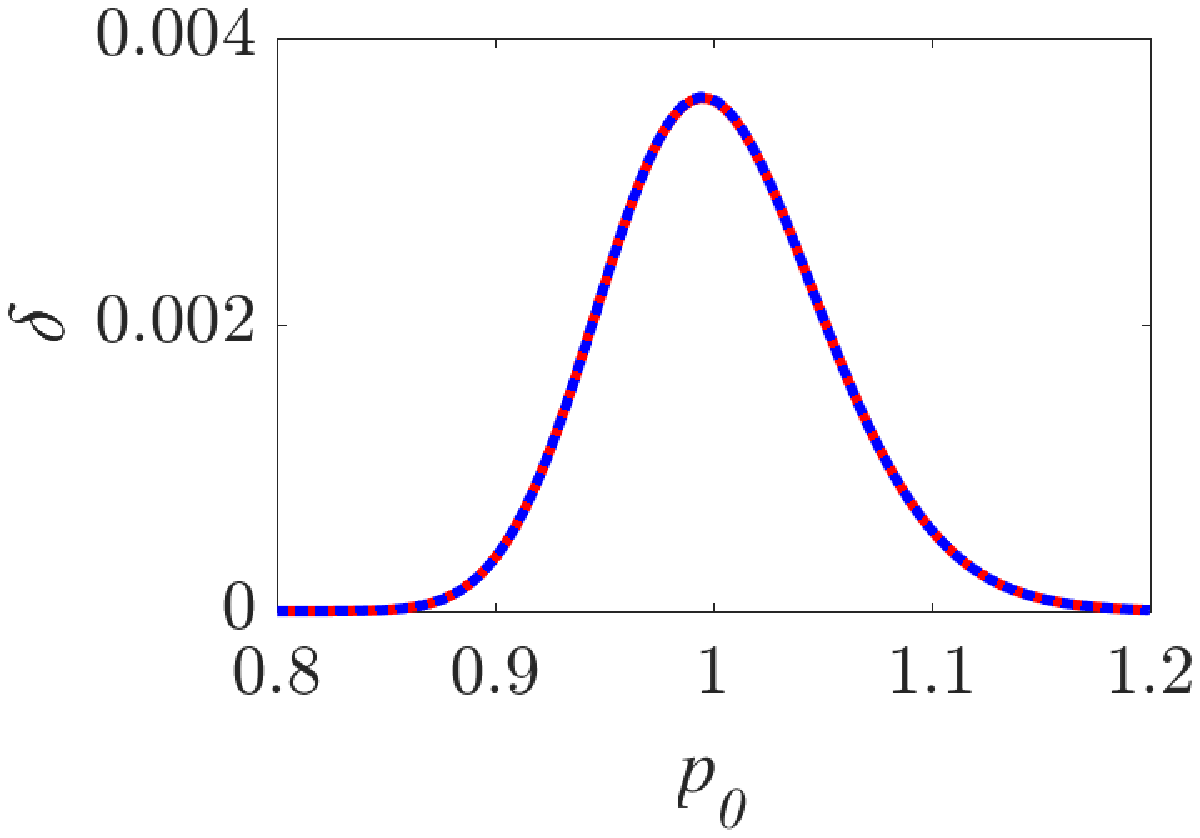}}
    \subfigure[]
        {\includegraphics[width=\scl\columnwidth]{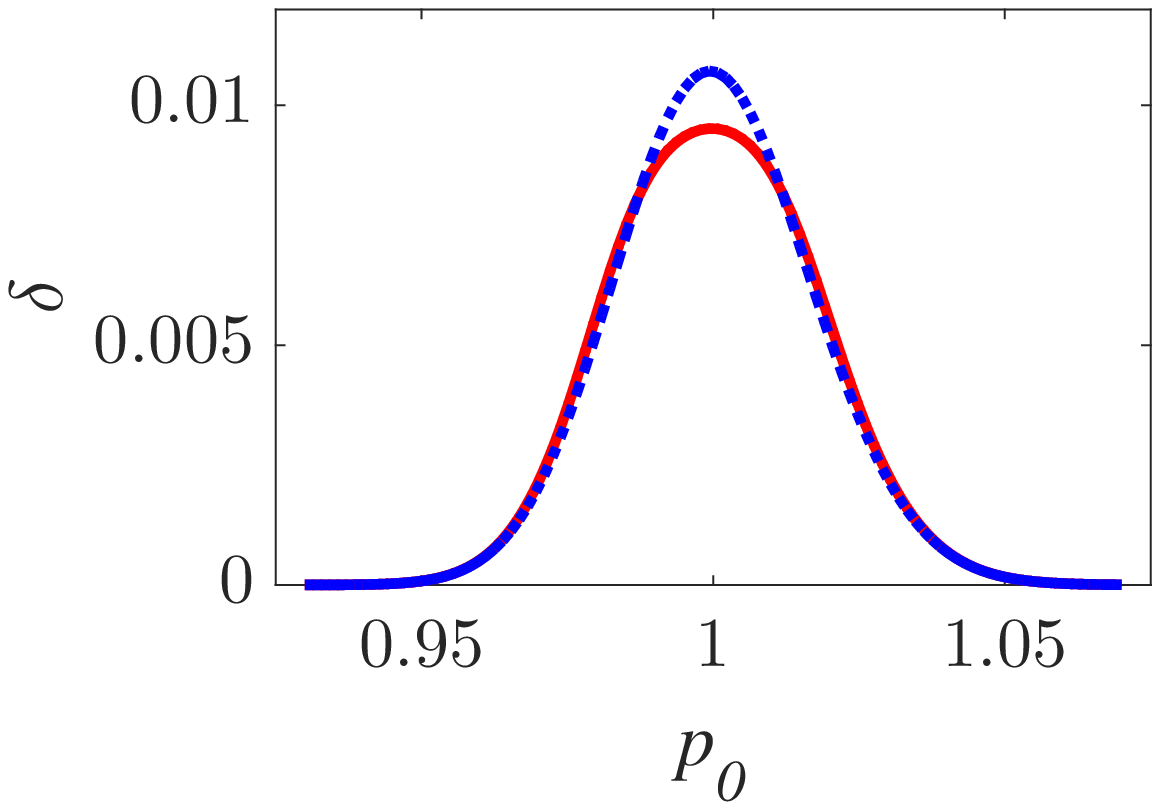}}
    \subfigure[]
        {\includegraphics[width=\scl\columnwidth]{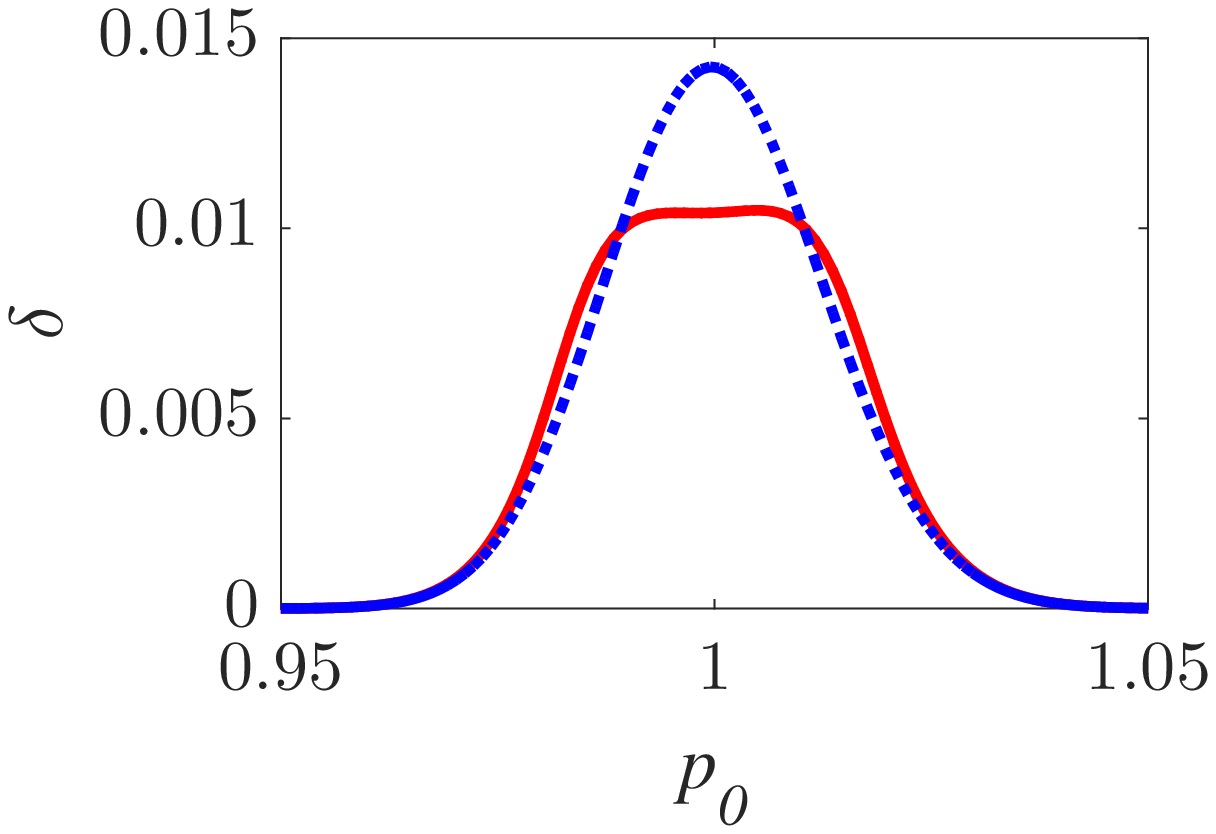}}
    \subfigure[]
        {\includegraphics[width=\scl\columnwidth]{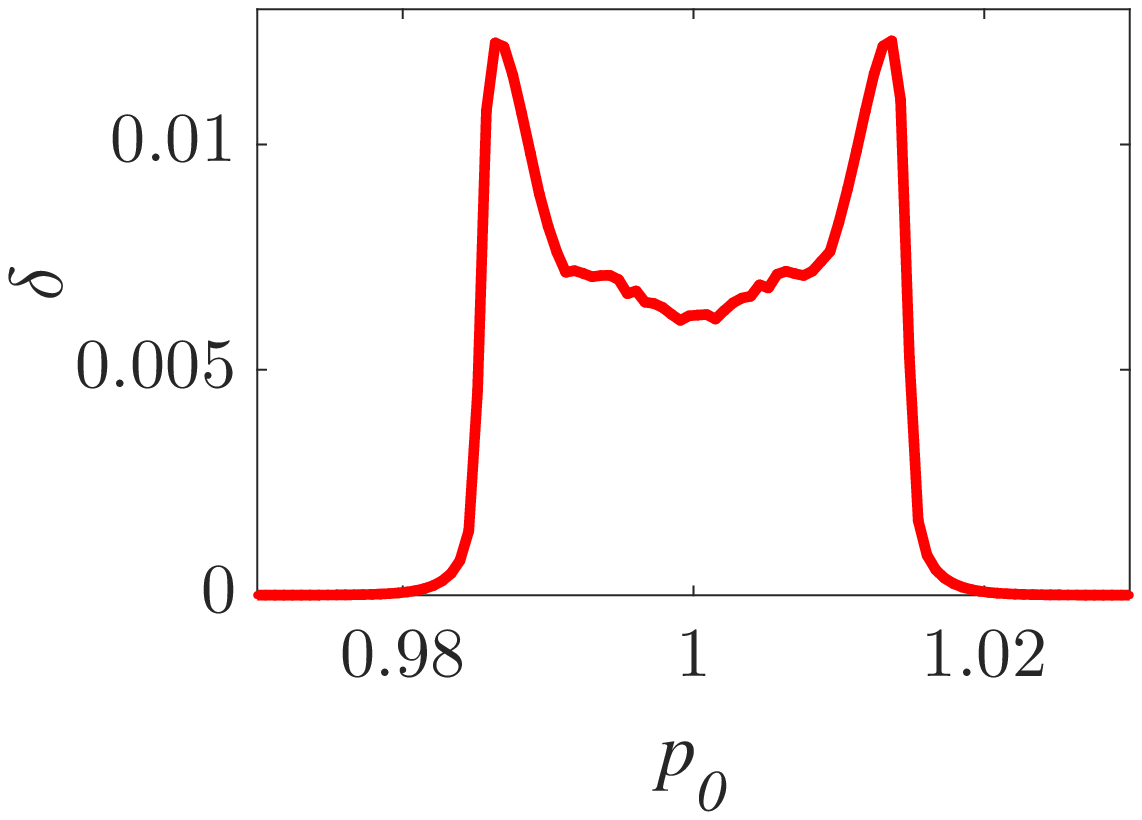}}
    \subfigure[]
        {\includegraphics[width=\scl\columnwidth]{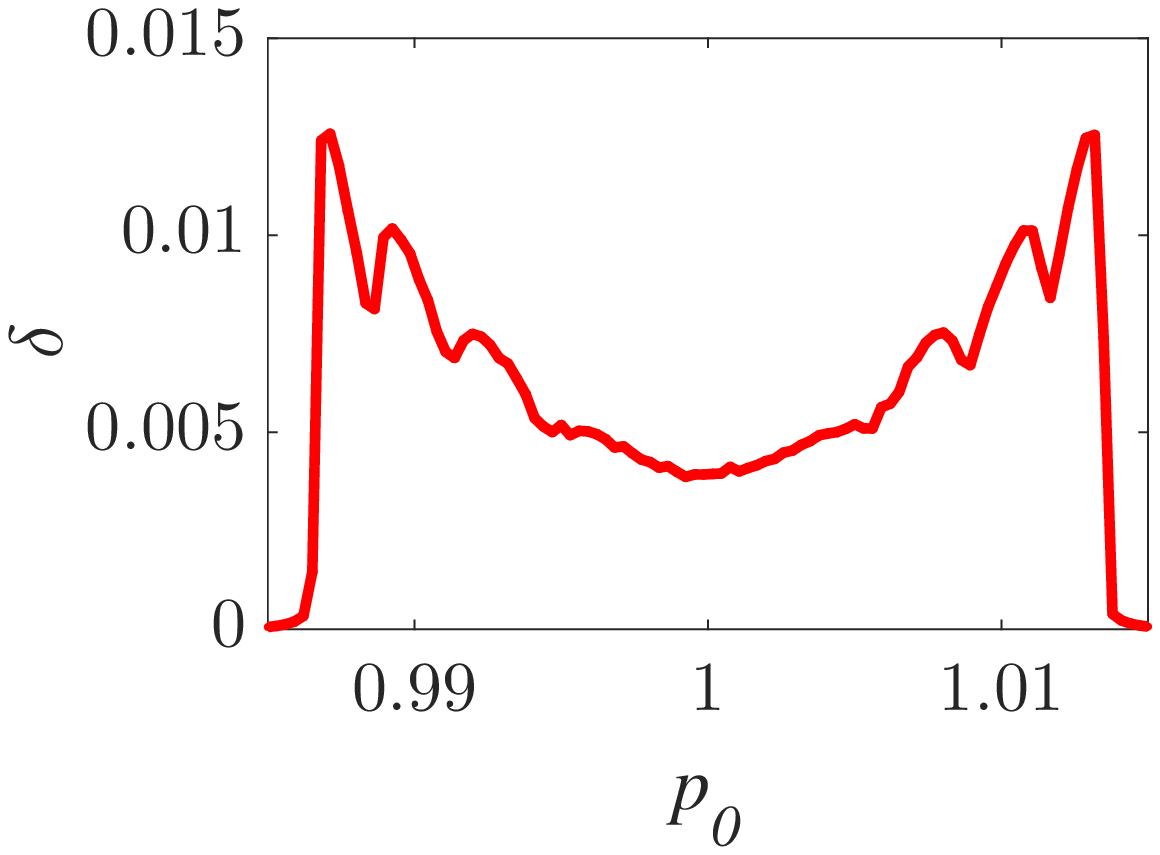}}
    \subfigure[]
        {\includegraphics[width=\scl\columnwidth]{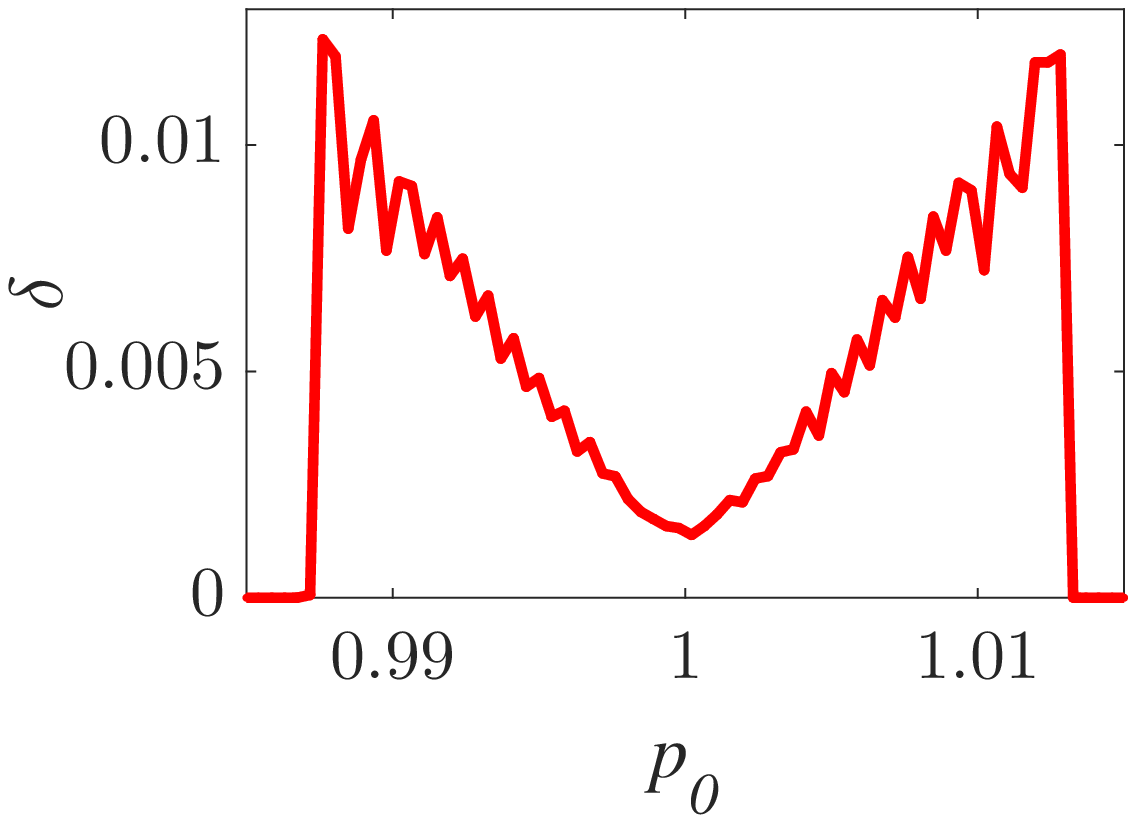}}
    \caption{Numerically (red, solid lines) and analytically (blue, dotted lines) calculated standard deviations $\delta$ of the momentum variations for an ensemble of particles with initial momentum $p_0$ and uniformly distributed initial positions after a transition through a wavepacket with relatively small amplitude $A=10^{-4}$, $v_p=1$, and various spatial widths $\sigma=20, 60, 80, 200, 400, 1500$ (a-f).}
\end{center}
\end{figure*}

\begin{figure*}\centering
\begin{center}
    \subfigure[]
        {\includegraphics[width=\scl\columnwidth]{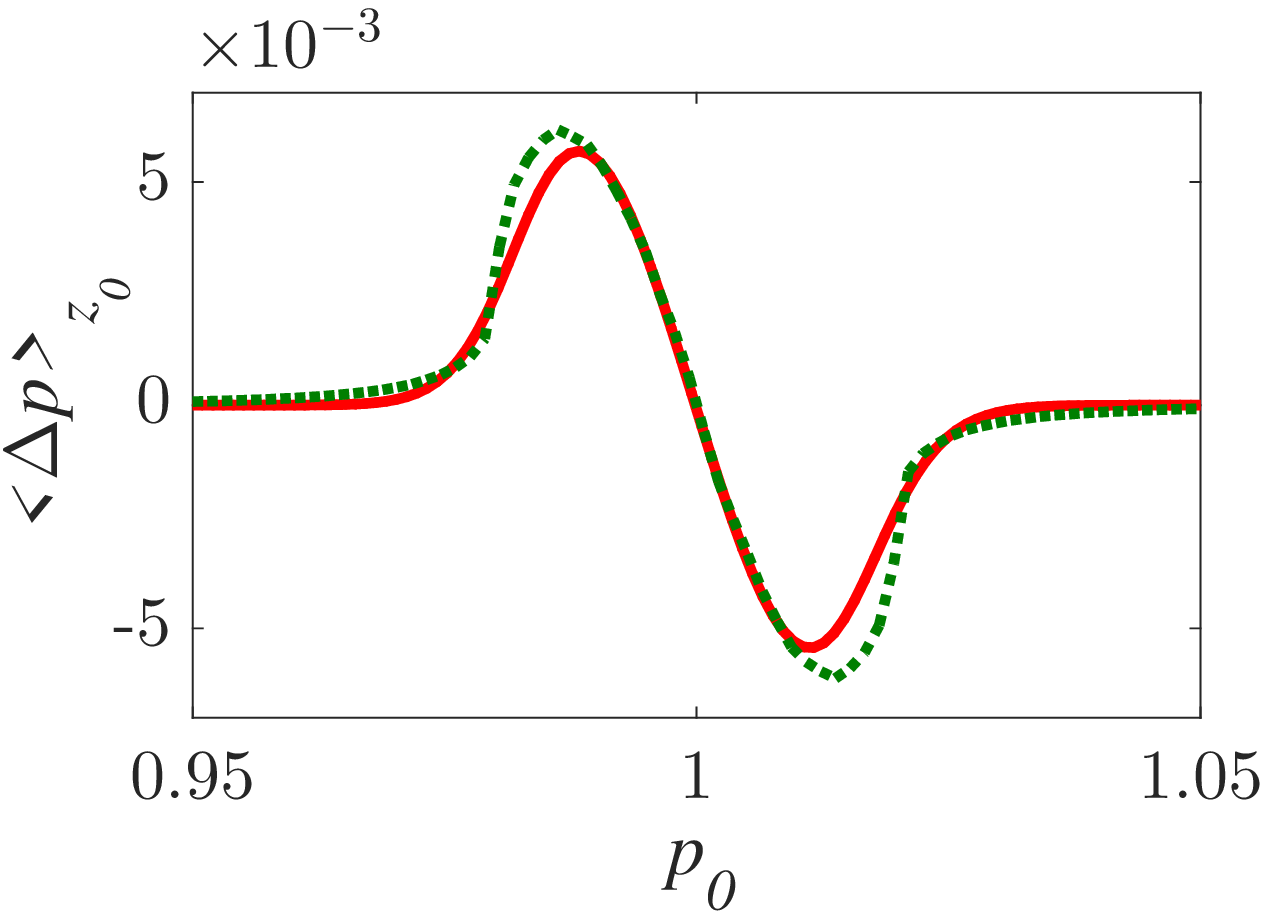}}
    \subfigure[]
        {\includegraphics[width=\scl\columnwidth]{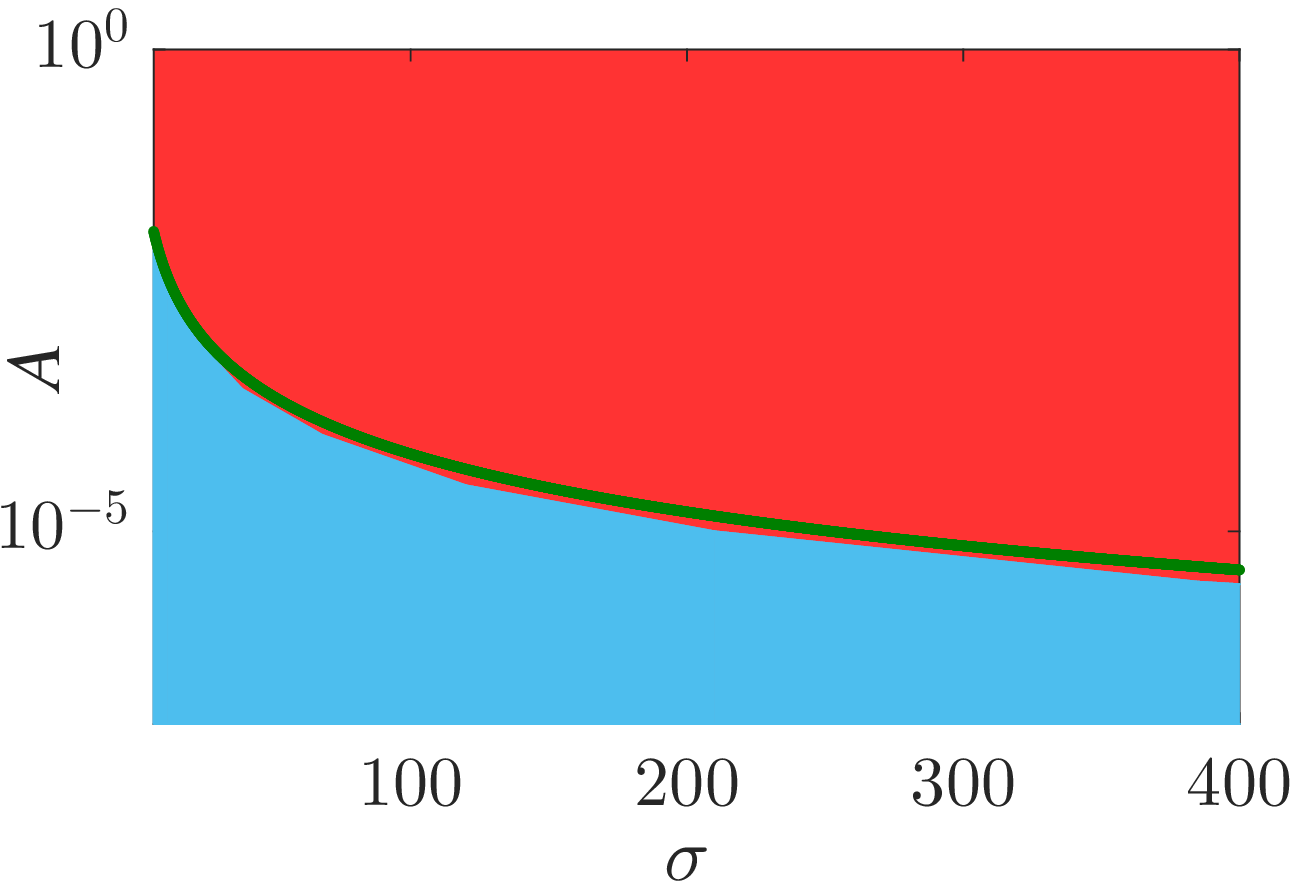}}
    \caption{(a) Numerically (red, solid line) and analytically (green, dotted line) calculated mean values of momentum variations $<\Delta p>_{z_0}$ for an ensemble of particles with initial momentum $p_0$ and uniformly distributed initial positions after a transition through a wavepacket with amplitude and width satisfying the condition (\ref{sigma_ON}), with the corresponding infinite-width wavepacket having $A=10^{-4}$. The analytical results are obtained from the analytical solutions of the pendulum-like Hamiltonian system (\ref{ONeil}). (b) Domain of validity of the analytical results obtained via the first-order canonical perturbation method in the $(A,\sigma)$ parameter space. The line corresponds to the condition (\ref{sigma_ON})}.
\end{center}
\end{figure*}

\begin{figure*}\centering
\begin{center}
    \subfigure[]
        {\includegraphics[width=\scl\columnwidth]{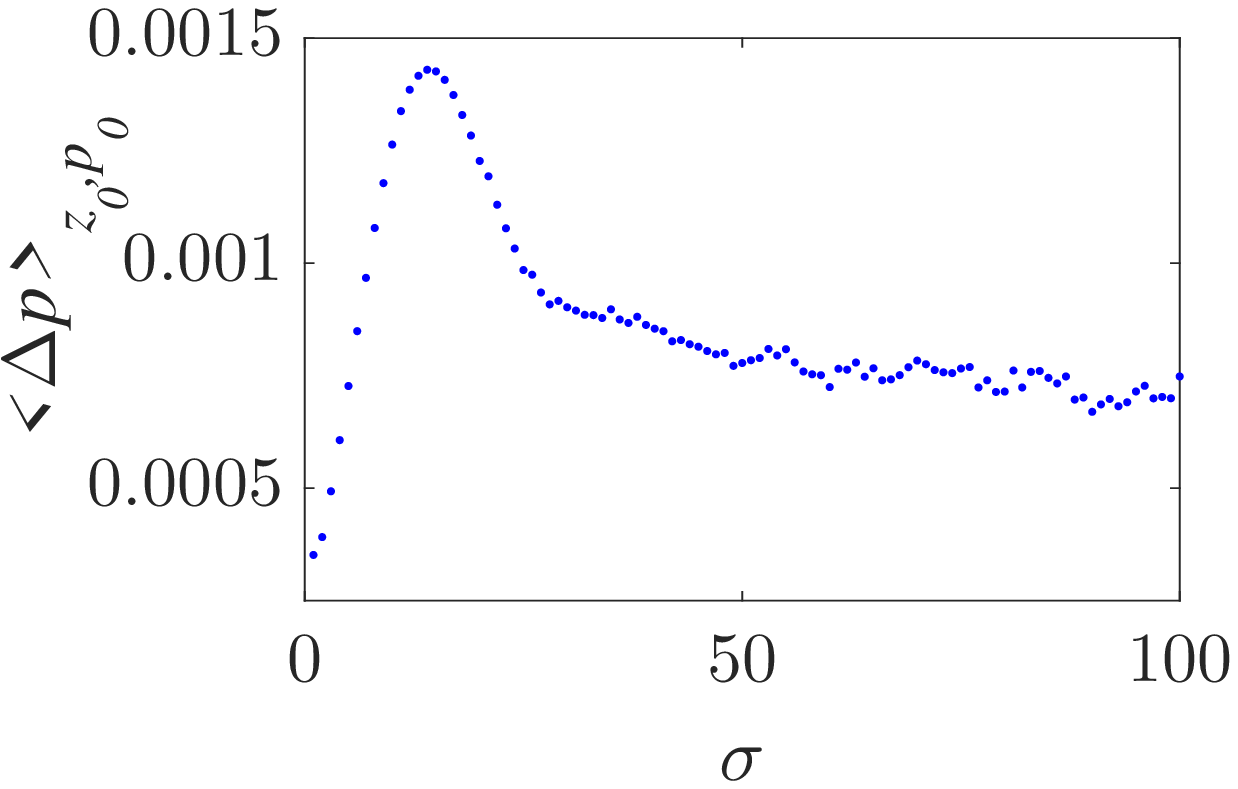}}
    \subfigure[]
        {\includegraphics[width=\scl\columnwidth]{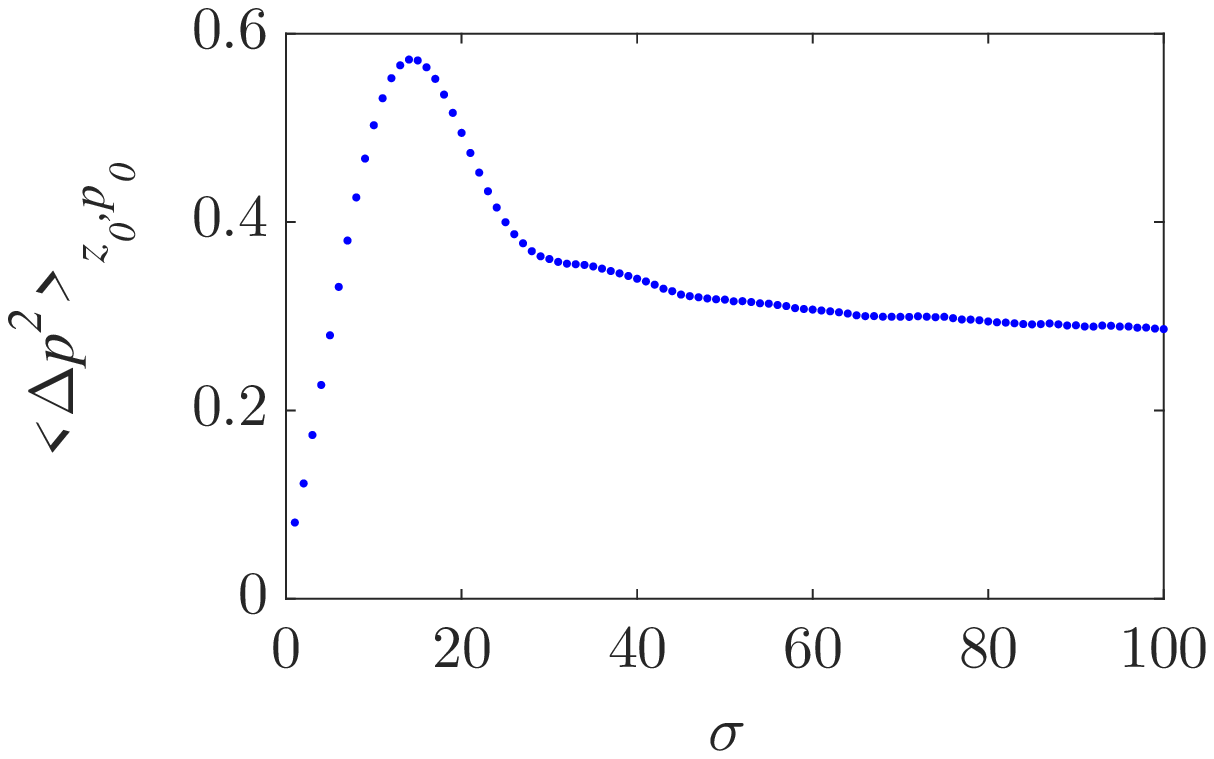}}
    \subfigure[]
        {\includegraphics[width=\scl\columnwidth]{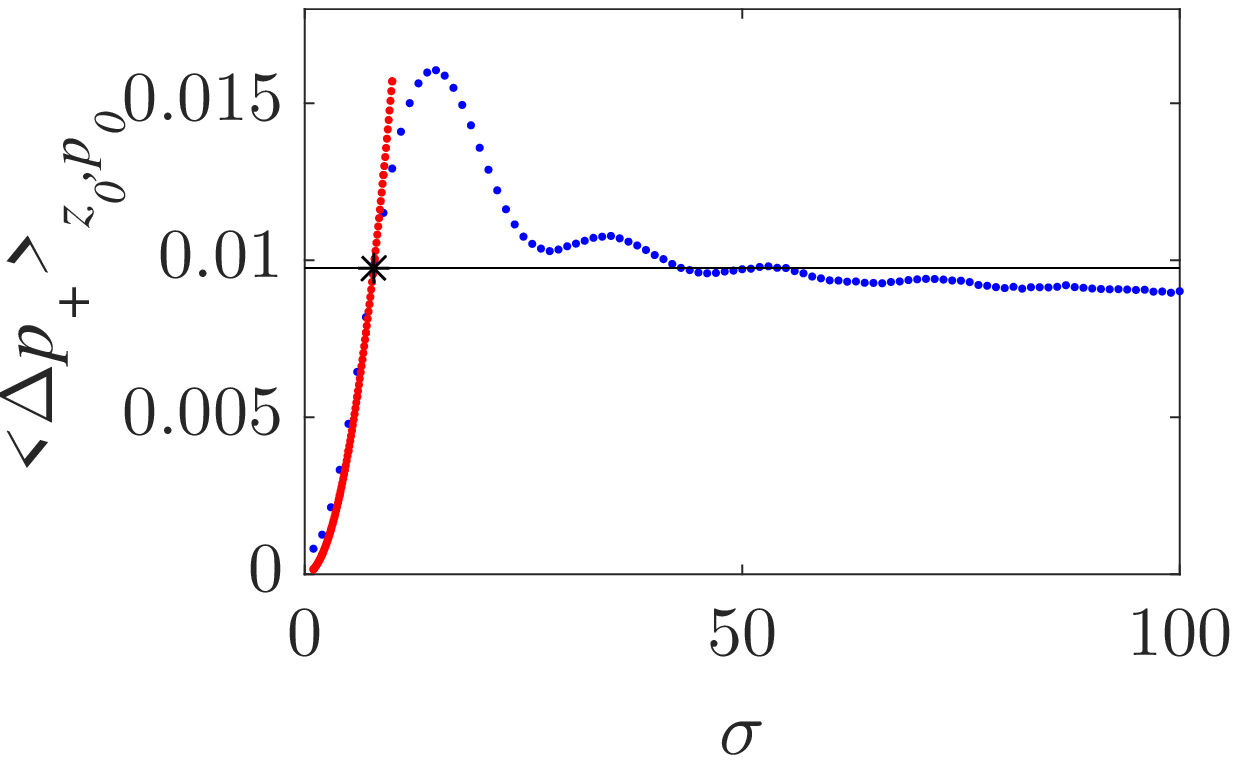}}
    \subfigure[]
        {\includegraphics[width=\scl\columnwidth]{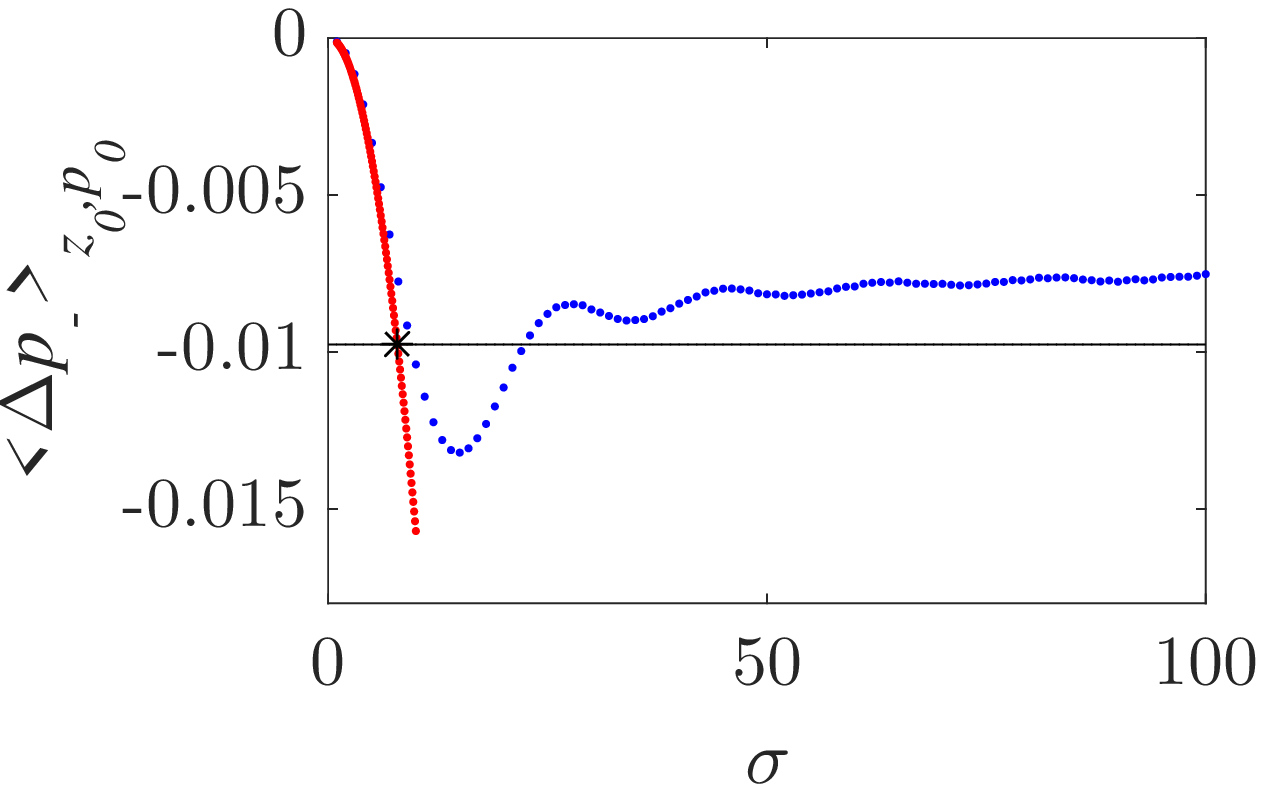}}
    \caption{Numerically calculated mean momentum $<\Delta p>_{z_0,p_0}$ (a) and energy $<\Delta p^2>_{z_0,p_0}$ (b) variation for an ensemble of particles with uniformly distributed initial positions and momenta, under interaction with a wavepacket with amplitude $A=10^{-2}$. The average momentum gain and loss $<\Delta p_\pm>_{z_0,p_0}$ are shown in (c) and (d), along with analytical curves and the point (asterisk) corresponding to the curve (\ref{sigma_ON}) in the $(A,\sigma)$ parameter space in Fig. 6(b).}
\end{center}
\end{figure*}

\begin{figure*}\centering
\begin{center}
    \subfigure[]
        {\includegraphics[width=\scl\columnwidth]{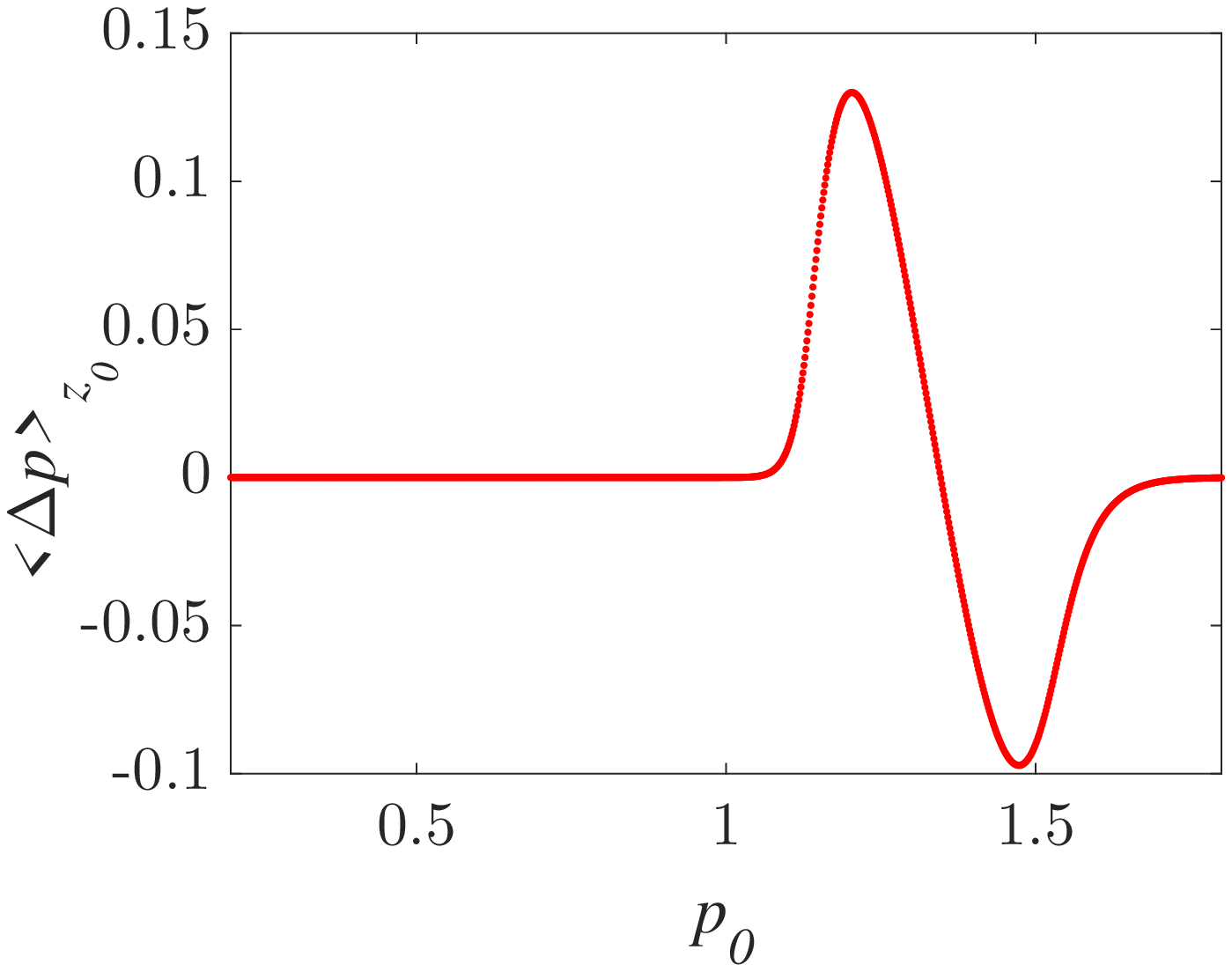}}
    \subfigure[]
        {\includegraphics[width=\scl\columnwidth]{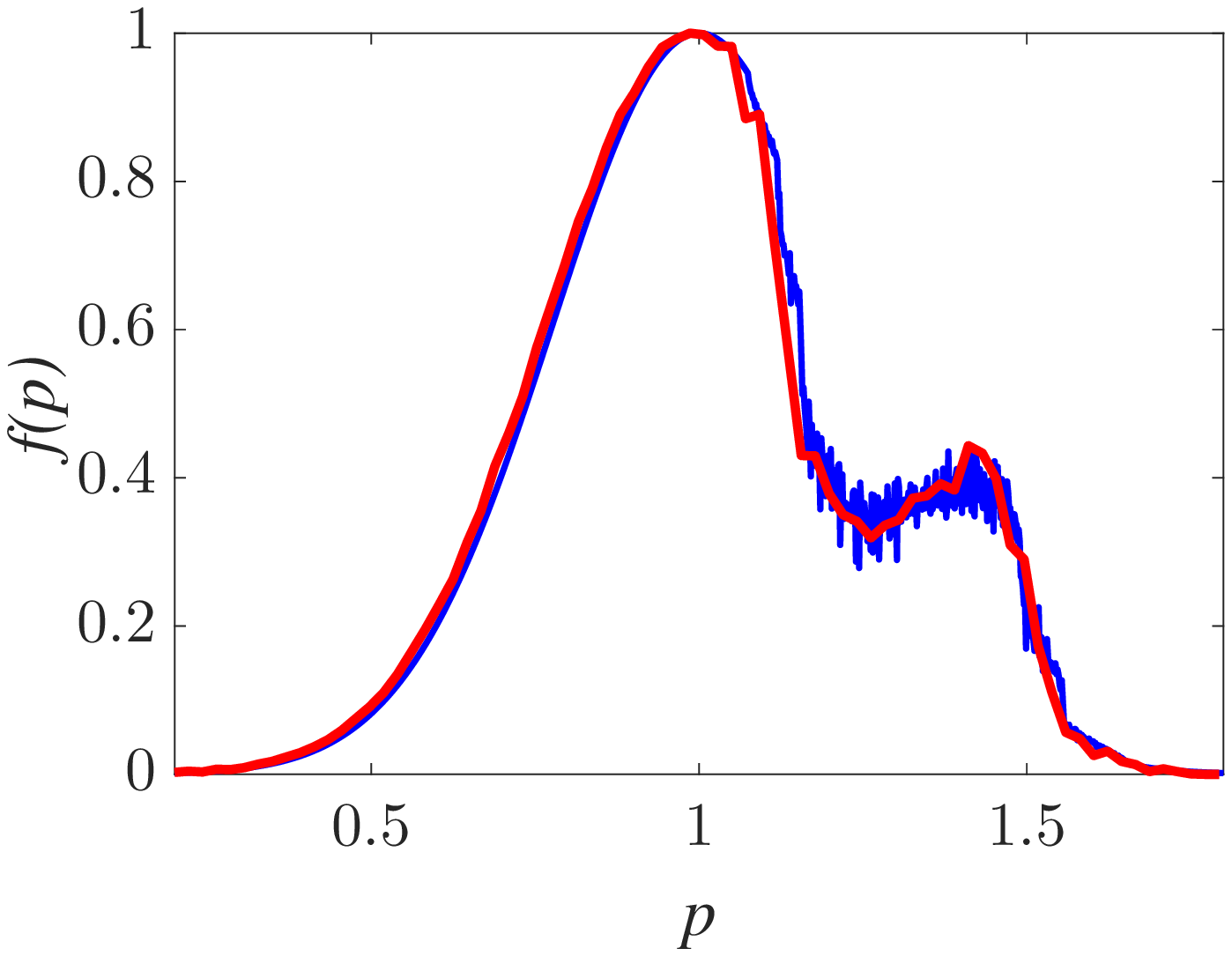}}
    \caption{(a) Mean momentum variation for an ensemble of particles with initial momentum $p_0$ and uniformly distributed initial positions after a transition through a wavepacket with $(A,\sigma)=(0.015,15)$ and $v_p=1.35$. (b) Final form of the initially Maxwellian momentum distribution function, as numerically calculated from a random Maxwellian momentum distribution (red line) and reconstructed from a weighted uniform initial momentum distribution (blue line) of $N=5\times10^4$ particles.}
\end{center}
\end{figure*}

\begin{figure*}\centering
\begin{center}
    \subfigure[]
        {\includegraphics[width=\scl\columnwidth]{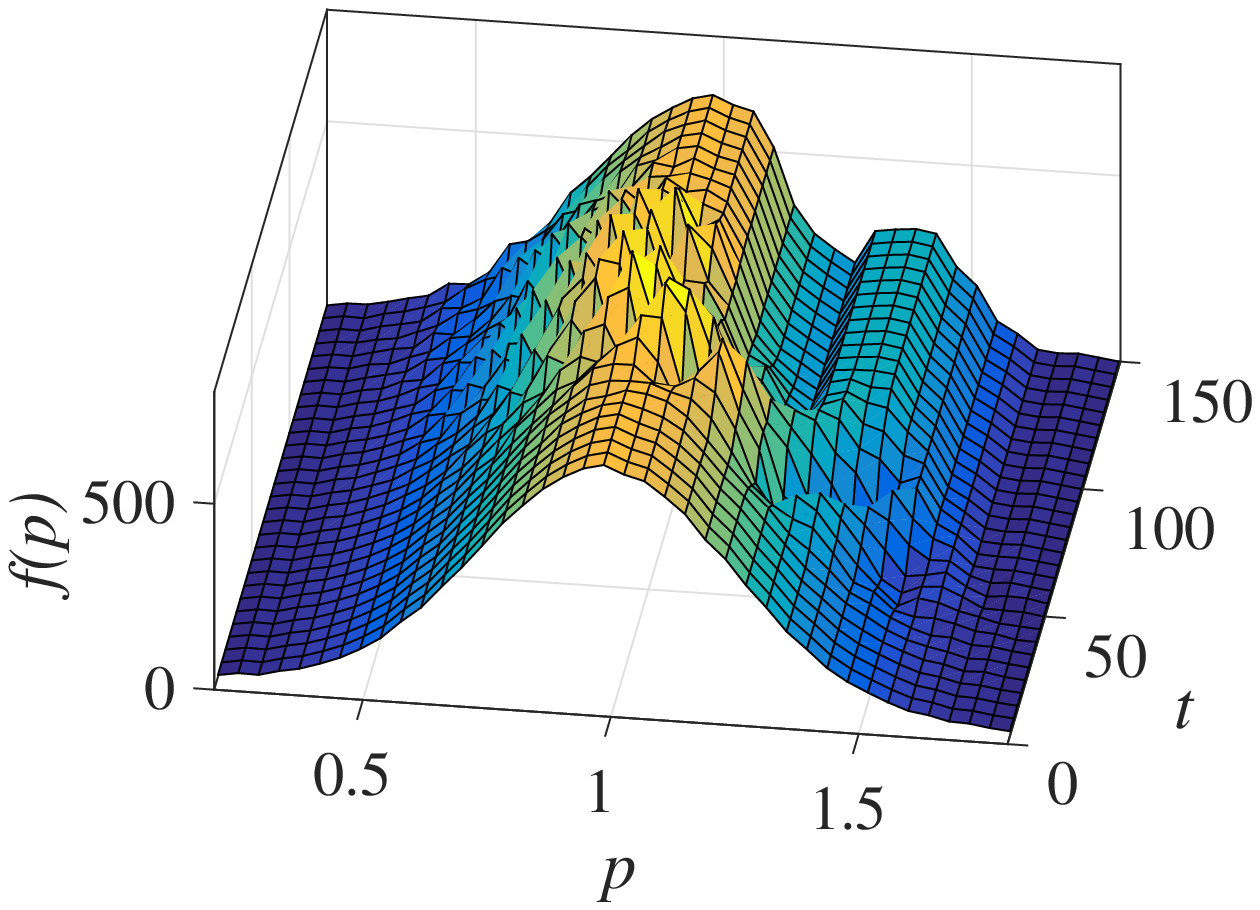}}
    \subfigure[]
        {\includegraphics[width=\scl\columnwidth]{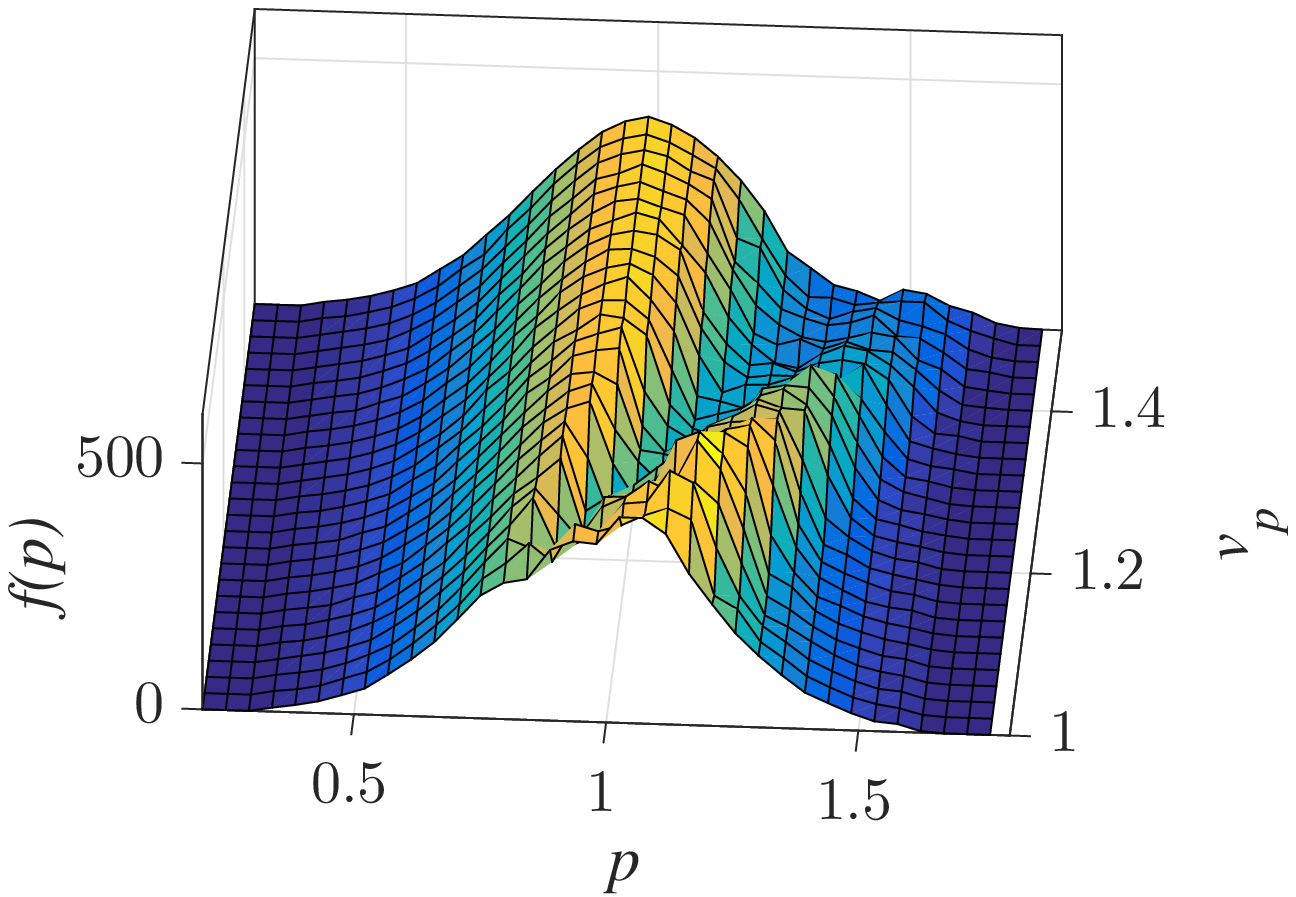}}
    \subfigure[]
        {\includegraphics[width=\scl\columnwidth]{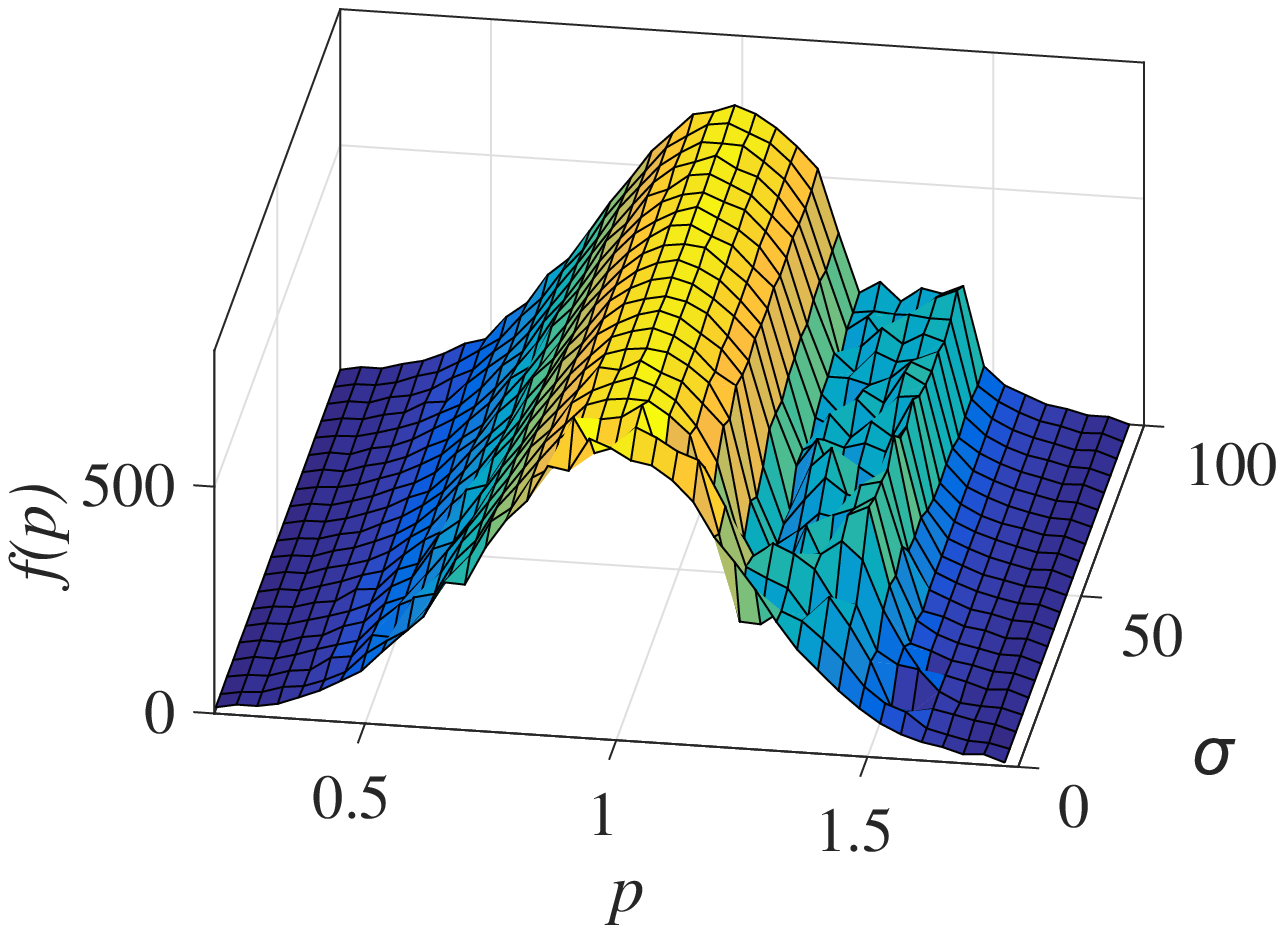}}
    \subfigure[]
        {\includegraphics[width=\scl\columnwidth]{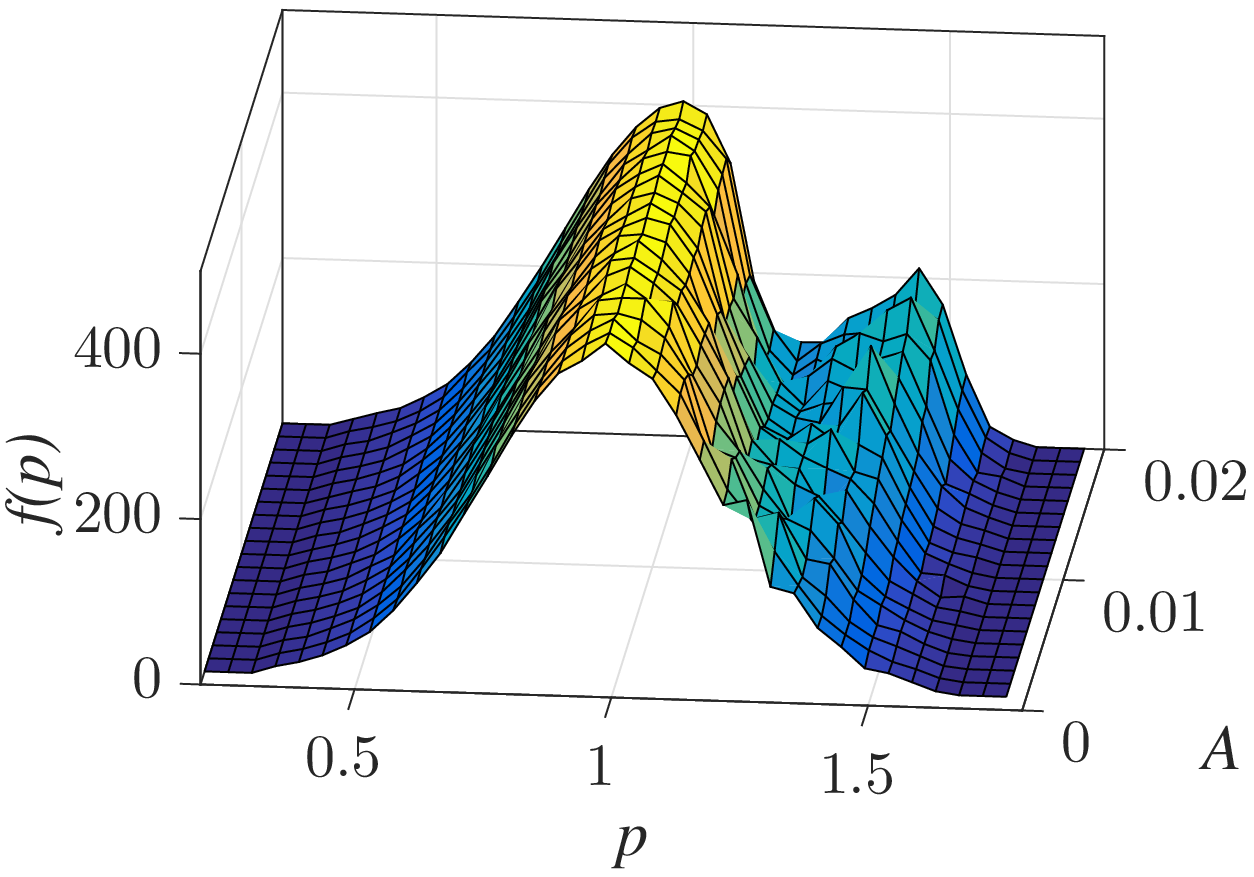}}
    \caption{(a) Time evolution of a Maxwellian momentum distribution function under interaction with a wavepacket having $A=0.015$, $\sigma=15$ and $v_p=1.35$. Dependence of the final form of the momentum distribution on the wavepacket phase velocity $v_p$ (b), width $\sigma$ (c), and amplitude $A$ (d), with two of the three parameters kept constant at the values of (a) and the other being varied. A random Maxwellian distribution of $N=10^4$ particles has been used.}
\end{center}
\end{figure*}

\end{document}